# Vision Transformers in Medical Computer Vision - A Contemplative Retrospection

Arshi Parvaiz[a], Muhammad Anwaar Khalid[a], Rukhsana Zafar[a], Huma Ameer[a], Muhammad Ali[a] and Muhammad Moazam Fraz[a,*]

[a]*School of Electrical Engineering and Computer Science, National University of Sciences and Technology (NUST), Islamabad, 44000, Pakistan*

ABSTRACT

Recent escalation in the field of computer vision underpins a huddle of algorithms with the magnificent potential to unravel the information contained within images. These computer vision algorithms are being practiced in medical image analysis and are transfiguring the perception and interpretation of Imaging data. Among these algorithms, Vision Transformers (ViTs) are evolved as one of the most contemporary and dominant architectures that are being used in the field of computer vision. These are immensely utilized by a plenty of researchers to perform new as well as former experiments. Here, in this article we investigate the intersection of Vision Transformers and Medical images and proffered an overview of various ViTs based frameworks that are being used by different researchers in order to decipher the obstacles in Medical Computer Vision. We surveyed the application of Vision transformers in different areas of medical computer vision such as image-based disease classification, anatomical structure segmentation, registration, region-based lesion Detection, captioning, report generation, reconstruction using multiple medical imaging modalities that greatly assist in medical diagnosis and hence treatment process. Along with this, we also demystify several imaging modalities used in Medical Computer Vision. Moreover, to get more insight and deeper understanding, self-attention mechanism of transformers is also explained briefly. Conclusively, we also put some light on available data sets, adopted methodology, their performance measures, challenges and their solutions in form of discussion. We hope that this review article will open future directions for researchers in medical computer vision.

## 1. Introduction

Advances in medical imaging modalities have made them indispensable in clinical practice. The analysis of these images by analysts is limited to human subjectivity, time constraints, and variation of interpretation, which leads to delusion [1, 2]. Medical images contain an ample information that is the key for medical diagnosis and hence treatment. The healthcare data comprises 90% of imaging data, so considered as the primary source for medical intervention and analysis. Multiple medical imaging modalities such as Computed Tomography (CT), ultrasound, X-ray radiography, MR Imaging (MRI), and pathology are commonly used for medical imaging diagnostics. Several challenging factors associated with medical imaging modalities such as expensive data acquisition [3], dense pixel resolution [4], lack of standard image acquisition techniques in terms of tool and scanning settings [5], modality-specific artefacts [6], hugely imbalanced data in negative and positive classes [7], sparse and noisy annotated datasets [8] are major hindrance in translating AI based diagnosis into clinical practice.

Since its surge, deep learning has shown remarkable success in automatic image analyses of medical imaging modalities. The advancements in deep learning have been flourished and perfected with time, revolving primarily around one algorithm called Convolutional Neural Networks (CNN). CNNs are potentially the most popular deep learning architecture for its distinguished capabilities to exploit the spatial and temporal relationship between the features of images, which need to be deciphered for extracting meaningful information hidden in images [9, 10, 11]. It has achieved notable accomplishment in medical imaging applications [12, 13] such as, determining the presence and then identifying the type of malignancy (Classification), locating the patient's lesion (Detection), extracting the desired object (organ) from a medical image (Segmentation), placing separate images in a common frame of reference

moazam.fraz@seecs.edu.pk (M.M. Fraz)
ORCID(s): 0000-0003-0495-463X (M.M. Fraz)

for comparing or integrating the information they contain (Registration), synthesizing images for balancing dataset (Generative Modeling) [14].

Despite that CNNs are very good at feature extraction tasks, they fail to encode the relative position of different features. In a CNN, the deeper layers are limited to view at whatever the initial layers have passed to them. This way they lose the global context of the features. Increasing the number of filters improves the representation capacity but at the cost of computation [15]. Various architectural changes are suggested by researchers for an efficient solution over the course of time and leading to attention mechanisms [16]. Using attention mechanism, the regions of the image are captured, to which the CNN should pay attention, and forwarded to the deeper layers. Researchers have demonstrated that replacing the convolutional layer with attention has improved performance [17, 18]. The breakthrough from Transformer network [16] in Natural Language Processing (NLP) tasks has inspired researchers to leverage this architecture for various computer vision tasks. Dosovitskiy et al. [19] proposed an adaptation to the transformer, known as Vision Transformer (ViT) that can be applied directly to sequences of image patches for extracting fine-grained features. In ViT, global attention is applied on 16x16 patches of the entire image, focusing on the global salient features of the image, which resolve the long-range dependency among image content. It gets the best out of the attention mechanism to incorporate global context in the image features without compromising computational efficiency.

The potential of the vision transformers is further explored by many researchers for solving various problems. However, in this survey we aim to highlight the contribution of vision transformers to circumvent the challenges in automatic diagnostic of diseases using medical imaging modalities and their applications in medical computer vision tasks. Our intended audience for this review are research practitioners from medical and interdisciplinary fields of computer vision. For their assistance, we have described commonly used terminologies and their description in table 1.

This review is organized into seven sections. Section 1 briefly discusses the role of deep learning, the emergence of the transformers, and the replacement of CNNs by transformers in medical computer vision. Section 2 discuss the organization and papers selection methodology and distribution of the Review. Section 3 discusses different medical imaging modalities and their application in the diagnostic and treatment of various diseases. Section 4 discusses the emergence of vision transformers and lists all the publicly available datasets used by the reviewed paper in every modality and deep learning task. Section 5 discusses visual recognition tasks to established the domain knowledge for the audience of the interdisciplinary field. Section 6 gives the details about the reviewed techniques categorized according to each deep learning task such as classification, segmentation, detection clinical report generation, and Miscellaneous which also include image registration. Section 7 identifies research gaps in the review papers and discusses the future directions for using transformers in medical computer vision.

## 1.1. Scope/Objective of the Review

The aim of writing this review paper is to highlight and discuss the contribution of Vision Transformers in medical computer vision across different medical imaging modalities. For this purpose, we have searched out papers from different top Conferences and Journals, excluding pre-prints, within the time span of three years from 2019-2022. The results achieved and the adopted methodology of each paper is reviewed comprehensively. The distinct categories that we reviewed belonging to the medical image analysis includes classification, detection, segmentation, registration, clinical report generation, image enhancement, image reconstructions and image synthesis. The literature of these categories is further divide into different medical imaging modalities. Ultimately, in addition to accentuating interesting techniques in the literature, we also put some light on research gaps and future directions. We hope this review will bridge the gap between computer vision community and medical specialists to foster the future research and development in medical computer vision This article is written keeping in mind the intended audience from the interdisciplinary fields, medical and AI. Publicly available datasets and downloadable links are listed in the table 3.

## 1.2. Comparison with other Reviews

Although there exists some reviews on transformers already which enfolds a significant amount of work, yet we feel that there is a lot of room for improvement. For example, no review is primarily focused on applications of vision transformers in medical domain. To bridge this gap, we come up with this survey in which our point of convergence is to scrutinize the exertion of ViTs in medical computer vision. To initiate this process we collected a bunch of articles addressing different transformers architecture and their utilization on multi modal medical images. We included almost 80 peer reviewed articles in our survey from prestigious platforms like PubMed, Springer, IEEE, Science Direct which makes our paper unique from other review articles. In addition to that we also explained different imaging modalities used in medical computer comprehensively. Moreover, we have given a brief note on available medical data sets in tabular form. The downloadable links to these data sets are also mention in these tables. We also discussed the results of state-of-the-art approaches in a well structured tabular form in which we described the performance metrices along with their results on the available datasets. In the end, we have also pointed out some challenges along with their insightful future directions. For comparative analysis of our review with Khan et al. [20] and Kai et al. [21] we visualized the main points in Figure 1.

**Table 1**
List of accronyms and abbreviations used in paper

| Acronyms | Words | Acronyms | Words |
|---|---|---|---|
| KID | Kernel & Inception Distance | MFTE | Multi-Branch Features Transformation and Extraction |
| AHA | Align Hierarchical Attention | MA | Microaneurysms |
| MSAM | Multi modal Spatial Attention Module | CLAM | Clustering-Constrained-Attention Multiple-Instance Learning |
| CAC | Coronary Artery Calcium | KFS | Key Factor Sampling |
| EMVT | Efficient Multi Scale Fusion Transformer | MSTGANet | Multi Scale Transformer Global Attention Network |
| NSCF | NonLocal Sparse Net Fusion | MCAT | Multimodal Co-Attention Transformer |
| TETRIS | Template Transformer Image Segmentation | FMNet | Feature Mapping Sub-Network |
| RDLs | Regionalized Dynamic Learners | CRC | Colorectal cancer |
| IDH | Isocitrate Dehydrogenase | MTTU | Multi Task Transformer Unet |
| VITBIS | Vision Transformer for Biomedical Image Segmentation | CIDEr | Consensus-based Image Description Evaluation |
| DAFNet | Disentangled Alligned and Fuse Net | ASFT | Adjacent Slices Feature Transformer |
| HYBRIDCTRM | Hybrid Convolutional Transformer | MTI | Multi Text Indexer |
| VIF | Visual Information Fidelity | PCR | Rpolymerase Chain Reaction |
| ABVS | Automated Breast Volume Scanner | PRCC | Papillary Renal Cell Carcinoma |
| FID | Frechet Inception Distance | GSM | Genitourinary syndrome of menopause |
| CEDT | Cross Encoder-Decoder Transformer | GLVE | Global-Local Visual Extractor |

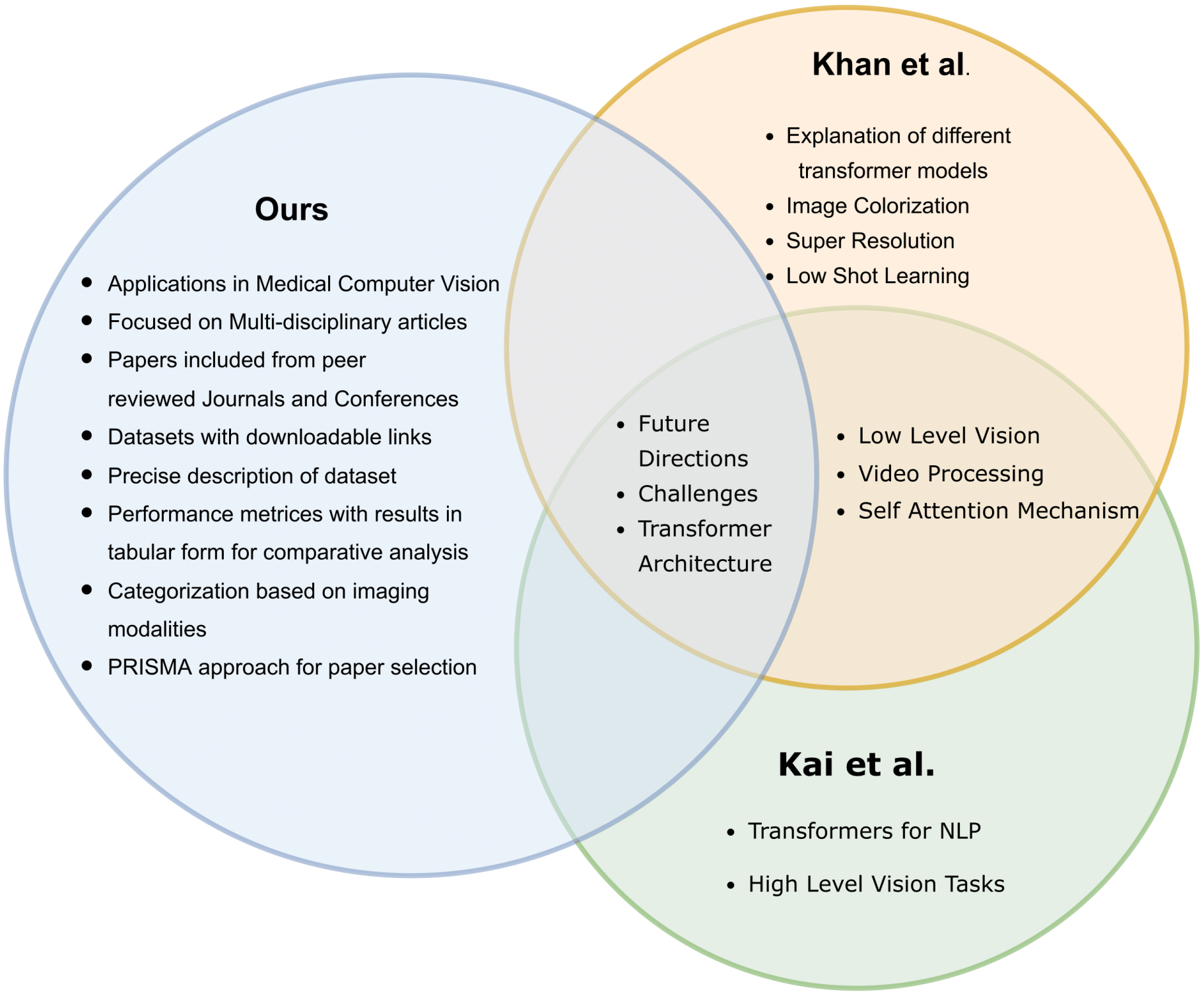

**Figure 1:** Comparison with recent published reviews on Vision Transformers

## 2. Survey Methodology

In this section we will discuss the study selection criteria based on which articles are chosen for the review, and distribution of the included articles according to venues (journals, conferences), medical imaging modalities, deep learning tasks (classification, segmentation, detection etc.) and impact factors.

### 2.1. Papers Selection

We have demonstrated the details of the searched and included research papers in this review article through PRISMA (Preferred Reporting Items for Systematic Reviews and Meta-Analyses). Figure 2 shows the summary of papers selection. We searched our papers on PubMed, Springer, IEEE Xplore, Science direct, and finally on google scholar. In the result of search queries, we have found 11060 papers. Among which 3060 were duplicate and were excluded from the study. We screened remaining 8000 and found 7,600 were not fulfilling the criteria of legitimacy for this survey as some of them was only about medical application without transformers and some of them was using transformer word in the different context than vision transformer. We further screened the remaining 400 articles and excluded the preprints from our study. In the PRISMA we have shown the categorization of our included 80 papers according to their application in medical domain. We have also demonstrated the distribution of the papers according to the medical imaging modalities.

### 2.2. Data Extraction Methods

We searched different platforms such as PubMed, Springer, Science Direct, IEEE Xplore and google scholar for extracting the research articles. We targeted top journals and conferences, in duration of last four years from 2019 to 2022. For extracting relevant papers for our study, we used different key words and combine them with the logical operators ‘AND’, ‘OR’ to get the better search results. The key words we used are:

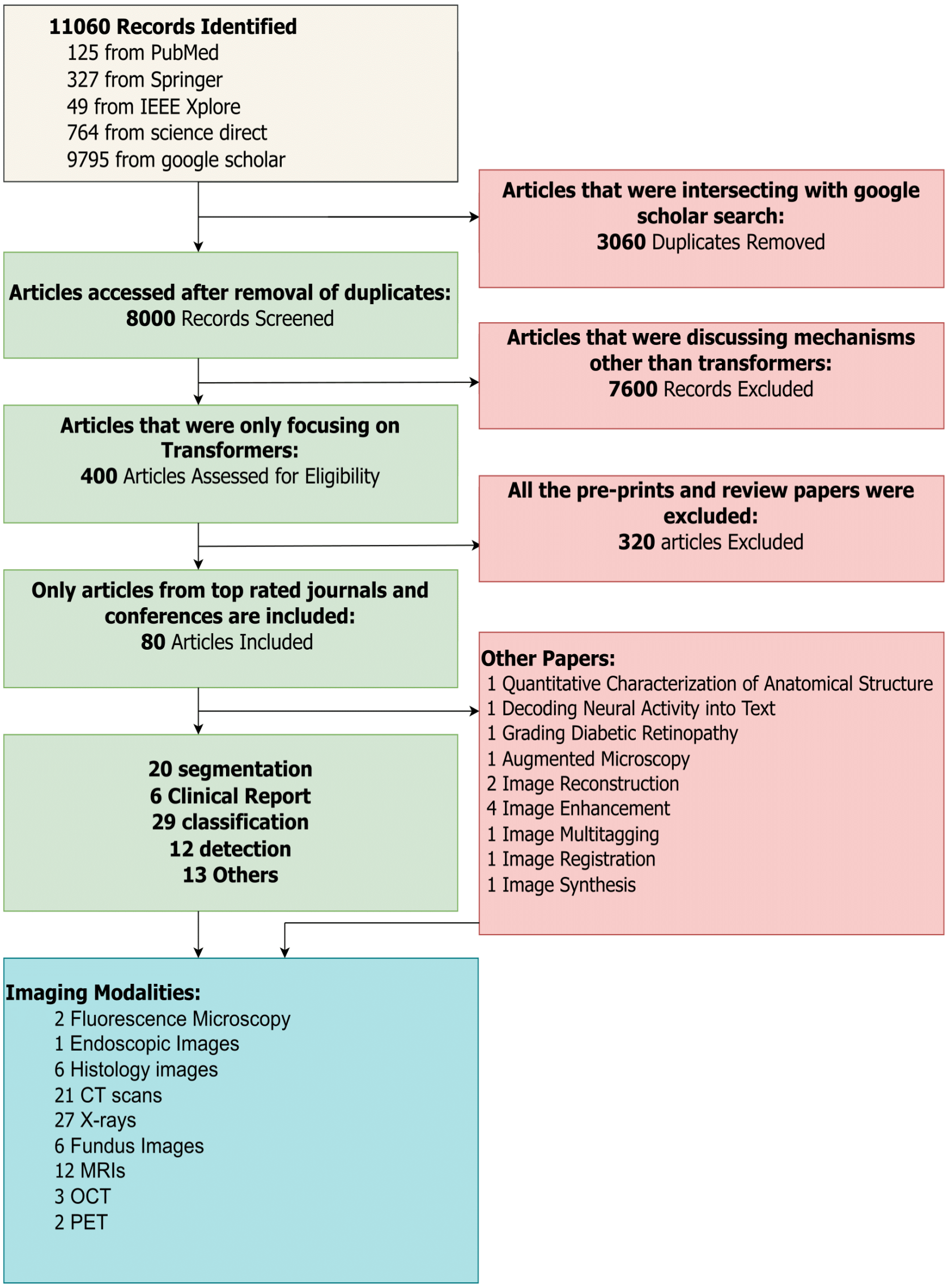

**Figure 2:** PRISMA - flow diagram of selected research articles for the review

- Computed tomography (CT) scans, Magnetic resonance images (MRI), Ultrasound, X-rays, Optical coherence tomography (OCT), Fundus images, Positron emission tomography-Computed tomography (PET-CT), Histopathology, Histology, WSI, Whole Slide Images
- Classification, Reconstruction, Segmentation, Registration, Detection, Report Generation, Enhancement
- Transformer, Vision Transformer

We extracted the keywords from all the articles on vision transformers included in our review and generate the tag cloud, which is shown in Figure 3. The tag cloud illustrates the trending terms in ViT applications in medical computer vision. As our central focus in this review is to recapitulate the application of vision transformers on medical imaging modalities, this word cloud mostly highlighting the applications (classification, segmentation, detection, denoising, captioning), medical imaging modalities (X-rays, PET, OCT, Whole-slide, CT, Histopathological, Fundus), disease (cancer, glaucoma, covid, diabetic, tumor, carcinoma), organs (retinal, chest, breast, pulmonary, brain) and other deep learning related terms like transformer, vision, attention, encoder, decoder, multi-model.

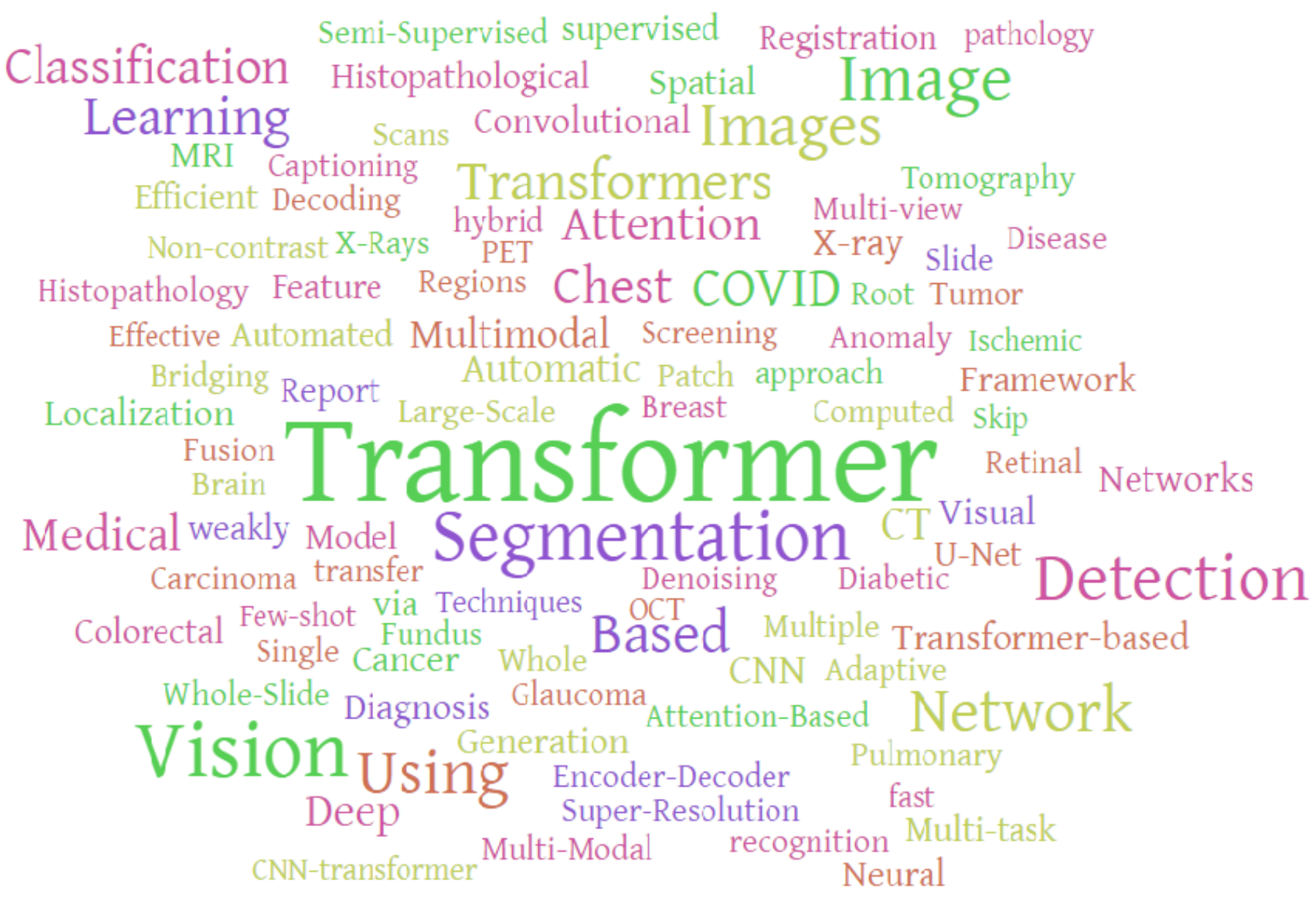

**Figure 3:** A visual depiction of most frequently used keywords in the reviewed articles

**Table 2**
Inclusion and exclusion criteria for papers selection

| Inclusion criteria | Exclusion criteria |
|---|---|
| Articles that address the medical computer vision tasks such as registration, segmentation, detection,classification, enactment, reconstruction and report generation using medical imaging modalities and vision transformer. | Articles which are not using medical imaging modalities and vision transformer. |
| Papers with proper evaluation metrics and detailed summary of proposed architecture including training parameters. | Articles that are not peer-reviewed. |
| Articles that are based on vision transformers. | Articles that are survey papers. |

The papers inclusion and exclusion criteria is given in table 2. Firstly, papers were selected on the basis of titles, if it does not match the inclusion and exclusion criteria then we read the abstract, conclusion, and model diagram for the final selection.

## 2.3. Papers Distribution

In this section we have shown the distribution of the published papers across journals, conferences, imaging modalities, impact factors, and medical computer vision tasks. The purpose of this section is to give the bird's eye view of the published work, that how much literature is available in top journals and conferences, what is the impact of the work, what imaging modalities are used and what is the progress of works across the years.

The Figure 4 shows the distribution of reviewed articles across the years. It can be seen in the graph that the literature for vision transformers has grown throughout the years from 2019 to 2022 as the applications of vision transformers in medical imaging modalities started growing from 2019 onward, with a large number of publications in the year 2021, in which more than 50 articles were published.

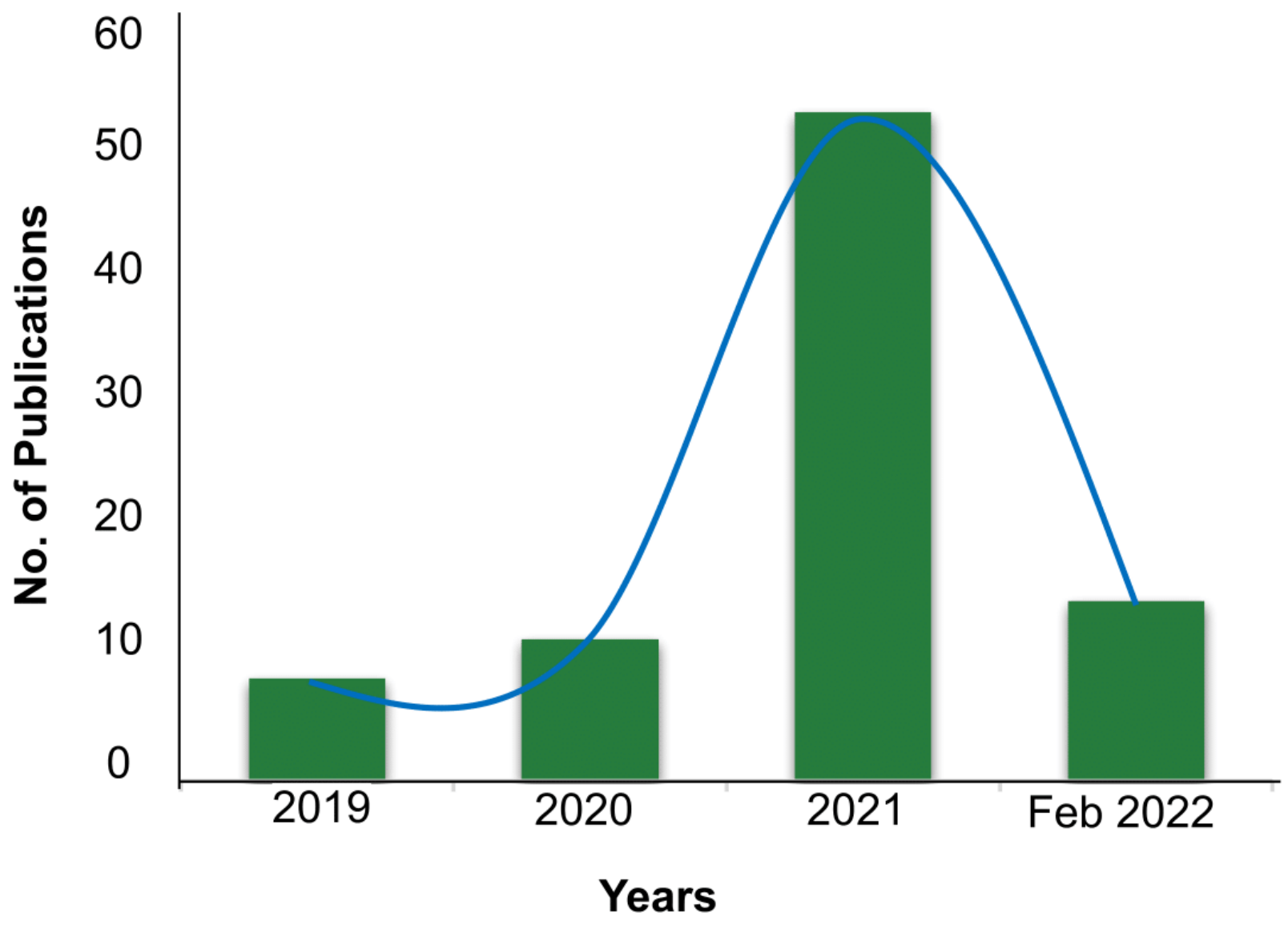

**Figure 4:** A chronological distribution of vision transformers research publications in medical image analytic.

Figure 5 shows the categorization of research articles based on visual recognition tasks using vision transformers. The recapitulation of the reviewed article is categorized based on the tasks such as classification, segmentation, detection, report generation, registration, and miscellaneous. Miscellaneous further contains different tasks such as reconstruction, enhancement and visual Neural Visual Content generation. The graph shows that most of the reviewed articles applied vision transformers on classification task which is 31%, segmentation 25%, Miscellaneous 19%, Detection 16%, Report generation 7%, and Registration 2%.

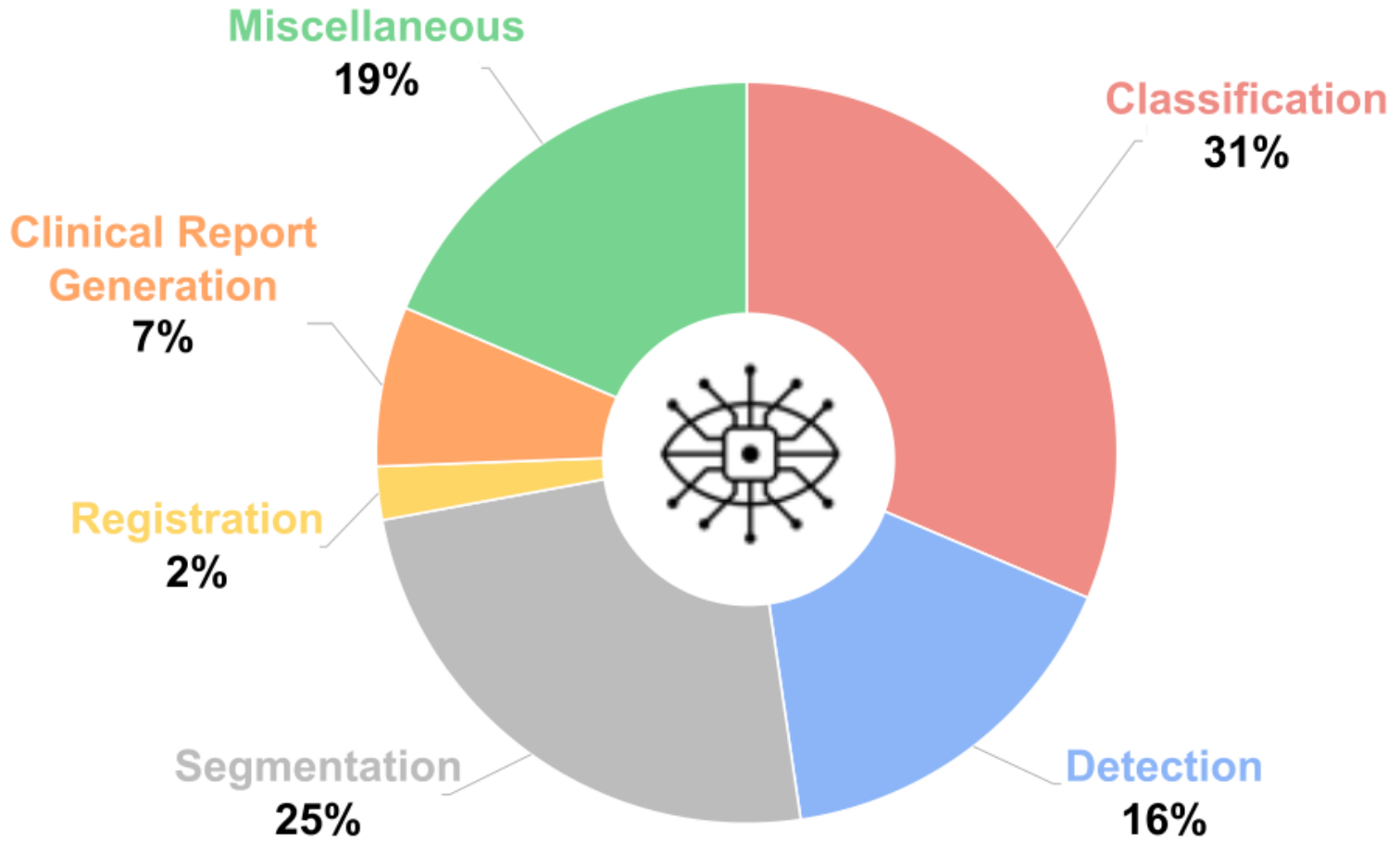

**Figure 5:** Distribution of reviewed articles based on visual recognition tasks.

As the review paper is focusing on the application of vision transformers in medical imaging modalities. Figure 6 depicts the statistics of imaging modalities used in our reviewed articles. Eight types of imaging modalities are used in this survey paper exploiting vision transformers include X-rays imaging modality which is 35%, CT Scans 26%, MRI Scans 13%, Histopathology Images 11%, OCT/Fundus Images 8%, PET 3%, Endoscopy 2%, and Microscopy 2%.

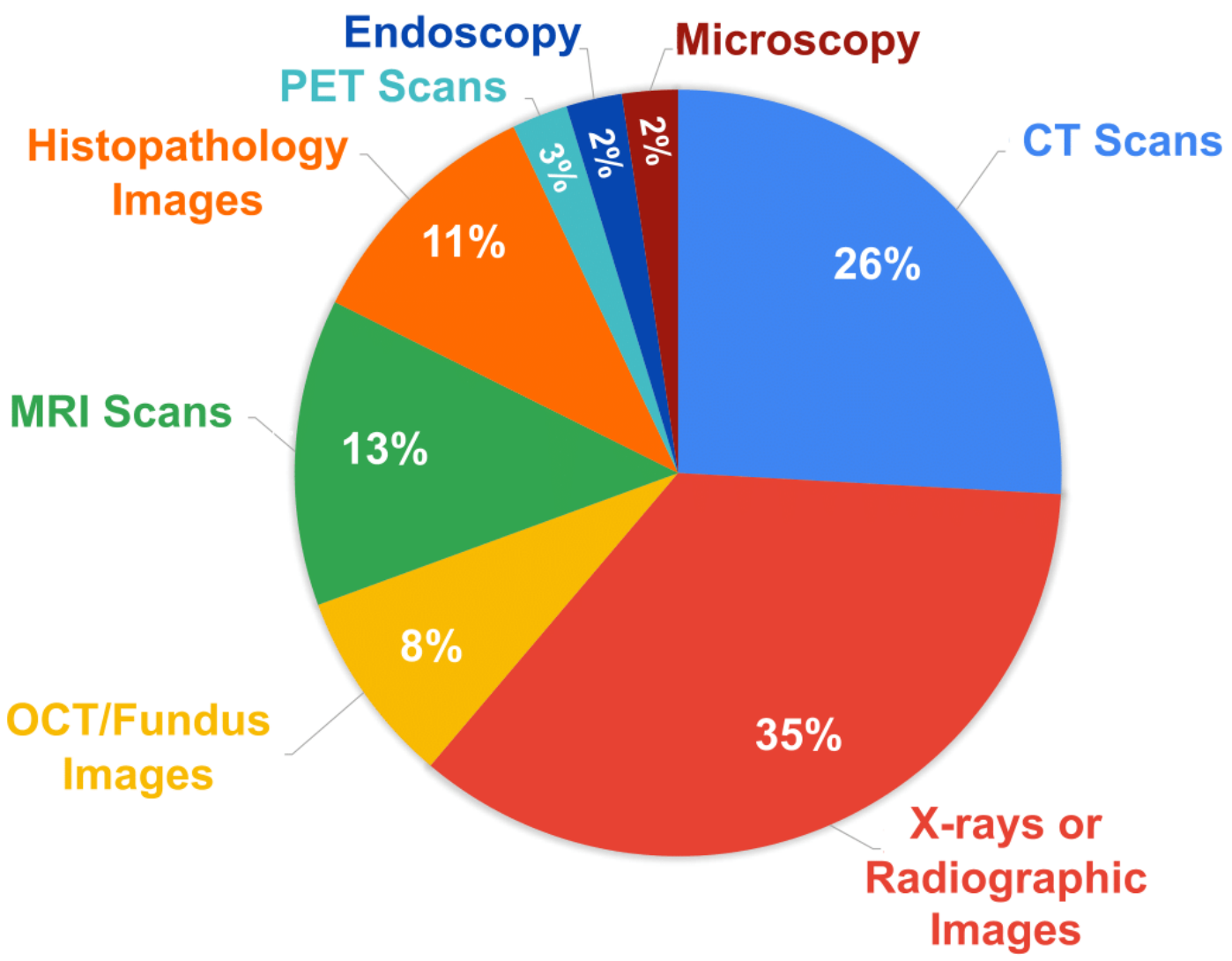

**Figure 6:** Dispensation graph of medical imaging modalities among the articles that are reviewed.

Figure 7 shows the count of vision transformer based medical imaging articles taken from various top journals and conferences. Each bubble represents the number of a specific journal or a specific conference and number of articles taken from these journals and conferences.

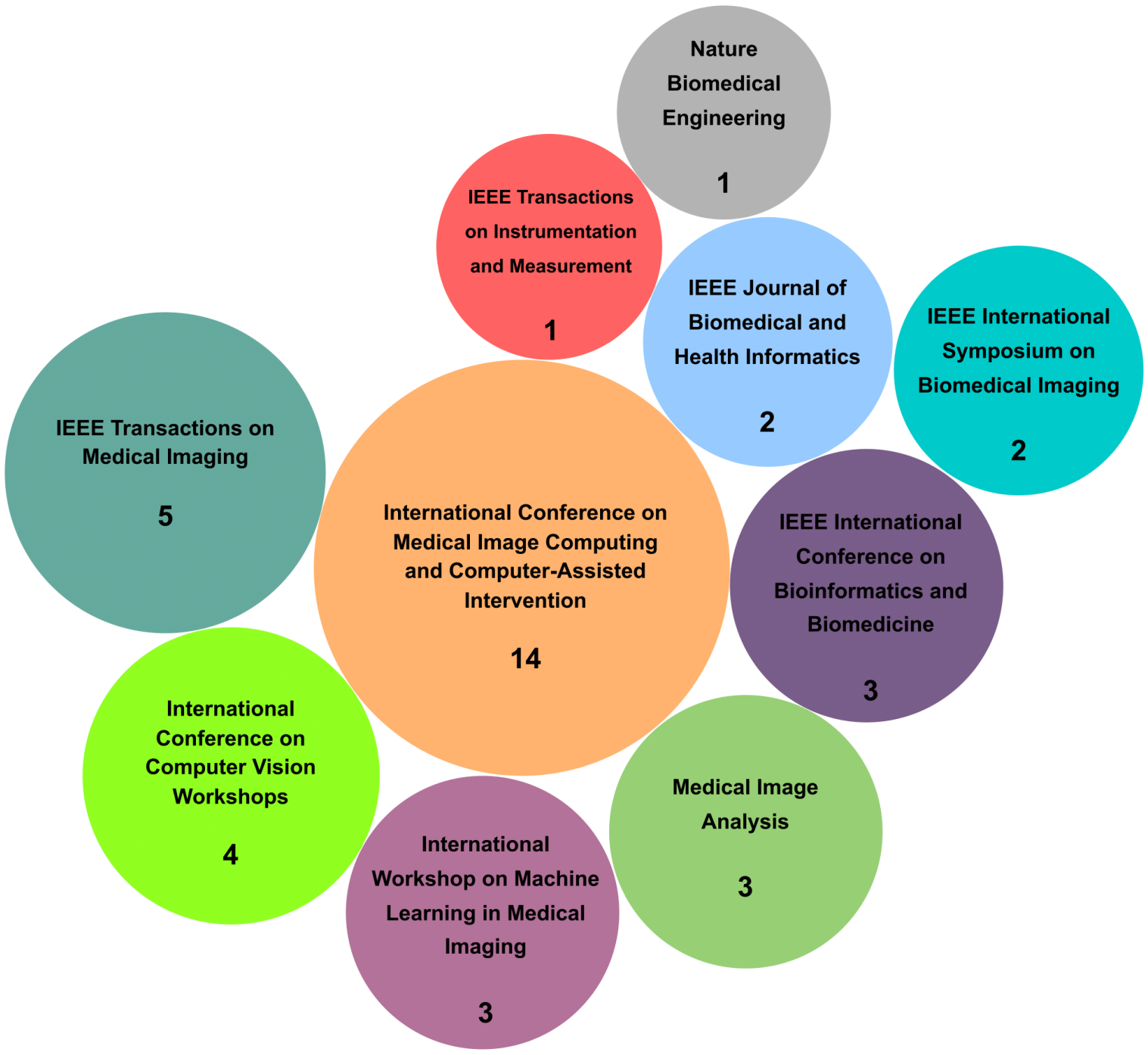

**Figure 7:** Bubble graph representing number of articles chosen from top ranked journals and conferences.

Figure 8 shows the distribution of vision transformer based reviewed papers across various journals of various impact factors according to JCR year 2020. The bubble size shows the number of reviewed articles retrieved from each journal. According to this figure the five papers are reviewed from IEEE Transaction on medical imaging with the impact factor of 10.048. The highest impact factor journal included in this survey paper is Nature Biomedical engineering which is 25.671.

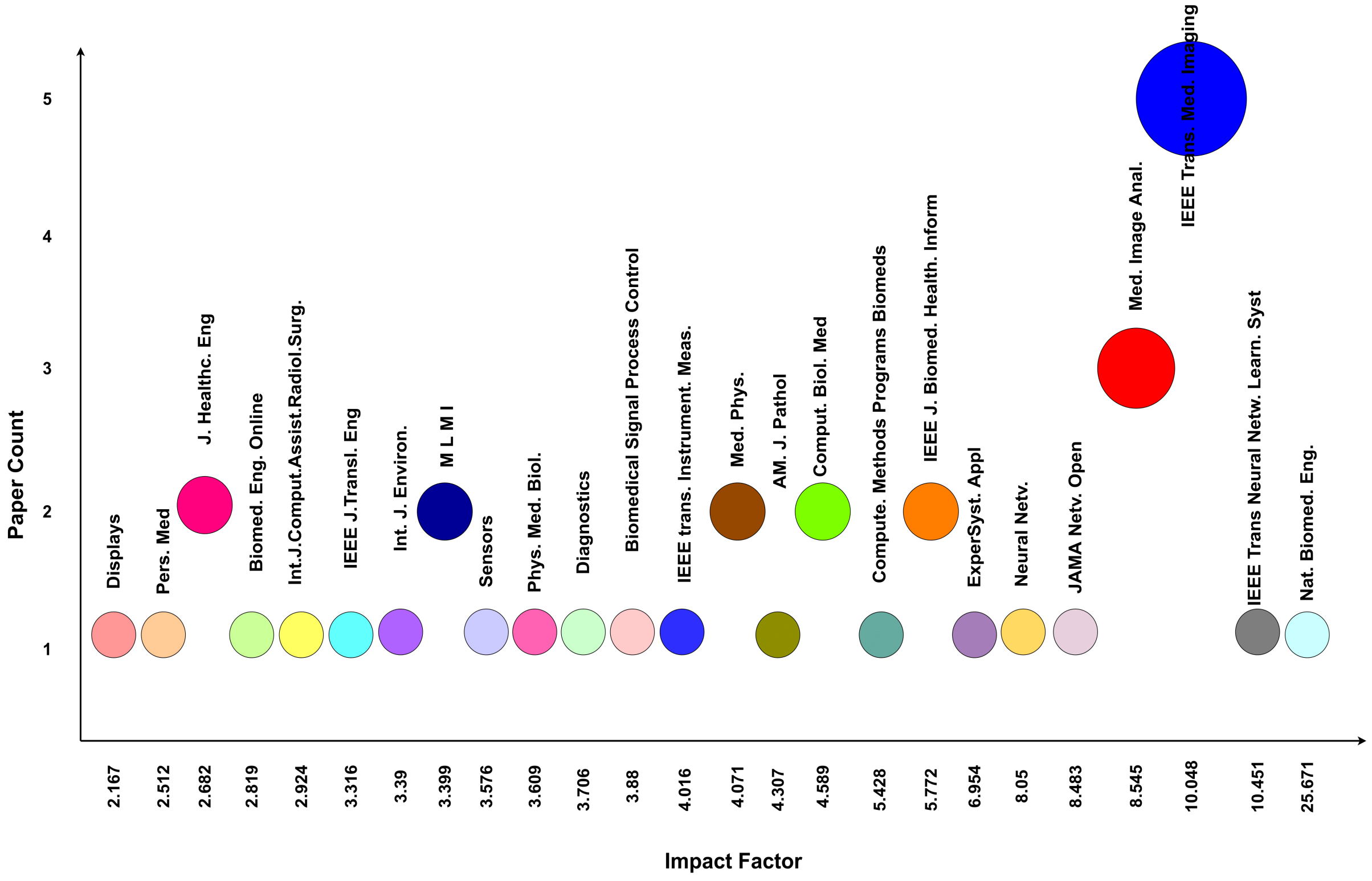

**Figure 8:** Visual representation of selected research publications from top tier journals along their impact factor.

## 3. A Delineation of Medical Imaging Modalities

This section is added for assisting computer vision practitioners to establish the basic domain knowledge of medical imaging modalities and their application in the diagnostic and treatment of various diseases. Medical images differ from natural images as they have specialized acquisition techniques. Physical phenomena such as electromagnetic radiation, sound, light, nuclear magnetic resonance, and radioactivity are used for generating medical images of external or internal organs of human body. These imaging techniques can be applied as non-invasive methods to view inside the human body, without any surgical intervention. Because of their importance in medical diagnostic a lot of advancement has taken place in image acquisition devices called image modalities. These image modalities play an important role in patients follow-up, regarding the growth of the already diagnosed disease state or undergoing a treatment procedure as 90% of the health data comprises of images. These imaging modalities are very crucial in public health and preventive measures as they help in establishing the accurate diagnosis. These medical images can capture different body regions such as eyes, chest, brain, heart, arms, and legs. There are different modalities of medical images such as computed tomography (CT), ultrasound, X-ray radiography, MR imaging (MRI), Positron emission tomography–computed tomography (PET-CT), pathology fundus images and Optical coherence tomography (OCT). The images acquired from these modalities are shown in Figure 9. Details about these image modalities are given in the subsequent section.

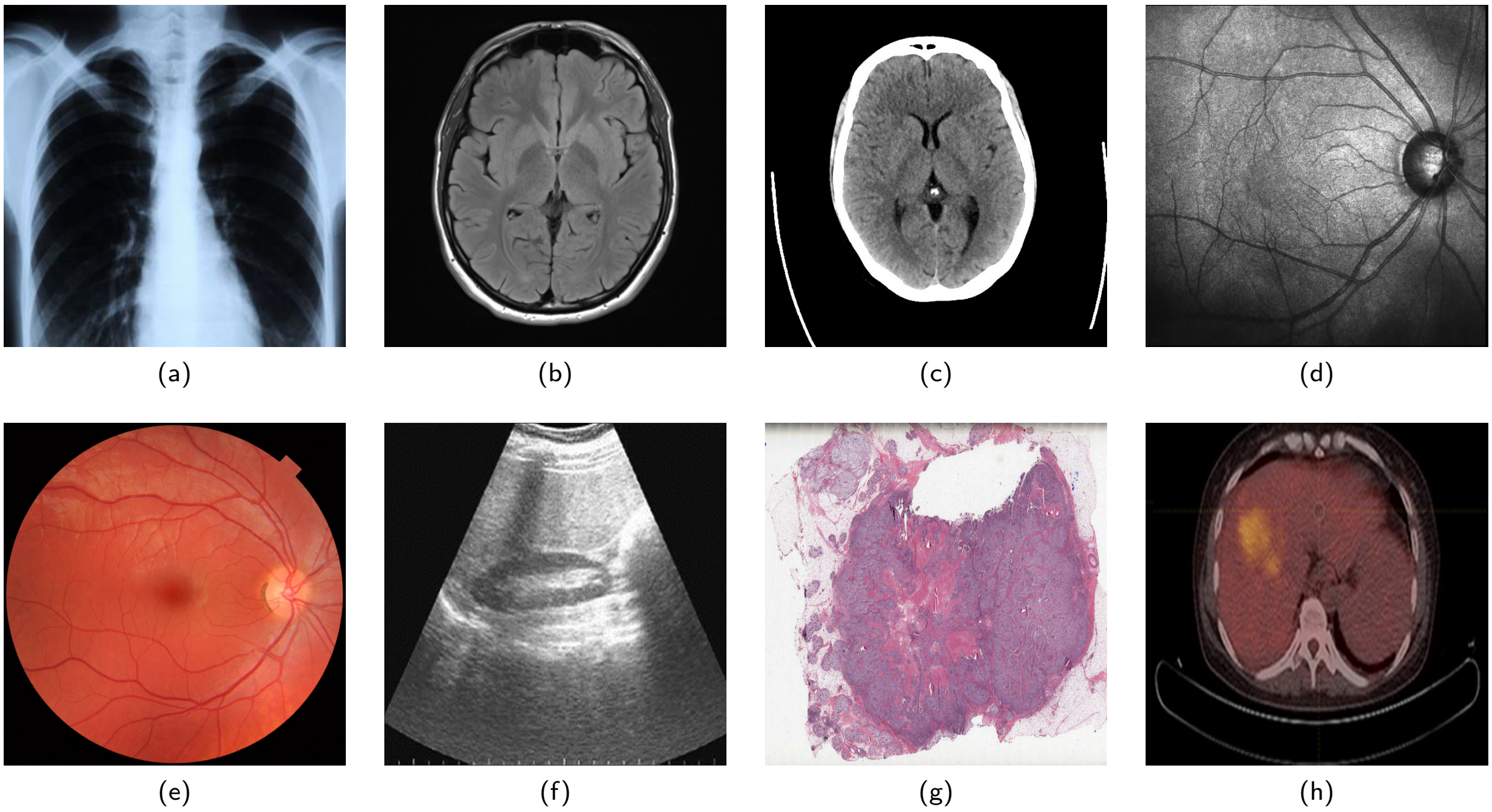

**Figure 9:** A catalogue of medical imaging modalities that vision transformers employed for assisted diagnosis. **(a)** Chest X-rays that are widely used for COVID-19 or pneumonia detection. **(b)** Brain MRI scans that are being used for diagnosis of aneurysms and tumors. **(c)** Brain CT scans that are employed to locate injuries, tumors, or clots leading to stroke. **(d)** OCT images that are playing an important role in diagnosis of retinal diseases such as age-related macular degeneration (AMD) and diabetic eye diseases. **(e)** Fundus images captured the rear of eye and are used for detection and grading of Hypertensive Retinopathy **(f)** Liver Ultrasound **(g)** Whole Slide Images (WSIs) that are being widely used in computational pathology **(h)** PET-CT scans that are responsible for detection and diagnosis of cancer, determining the spread or recurrence of cancer or metastasis

### 3.1. X-ray Imaging

According to National Institute of Health (NIH), US [22], X-rays images are captured non-invasively using radiation that is part of the electromagnetic spectrum. X-rays are mostly captured for diagnosing bone fracture [23], but chest x-rays are also used for detecting pneumonia [24]. X-rays are also used by mammograms for breast cancer detection [25]. Other most familiar uses of X-rays are for breast tumors [26], enlarged heart [27], blocked blood vessels [28], conditions affecting your lungs [29] , infections [30], osteoporosis [31], arthritis [32], tooth decay [33].

### 3.2. Computed Tomography (CT) Scans

National Institute of Health (NIH), US [22] described computed tomography (CT) scan is a computerized x-ray imaging technique in which a narrow beam of radiation is focused and then quickly rotated around the body to capture the detailed internal images, called tomographic images, of the body's slice non-invasively. CT Scan produces 2-dimesional as well 3-dimensional images of slice of the body. Once several images are taken these images are digitally stacked together to form three-dimensional images. CT Scans are used for identifying the various organs/slices of the body for example CT scan of the heart is used for detecting various types of heart disease or abnormalities [34]. CT Scans of the head, to locate injuries [35], tumors [36], clots leading to stroke, hemorrhage, and other conditions [37]. CT Scans of the lungs is used for detecting cancer [38], tumors excess fluid, pulmonary embolisms (blood clots) [39],lung infections [40] and emphysema or pneumonia [41].

### 3.3. Optical Coherence Tomography (OCT) & Fundus Images

According to American Academy of Ophthalmology (AOA) [42], Optical coherence tomography (OCT) captures invasive cross-section images of the retina using light waves. OCT can be used to examine the retina's distinctive layers which help in mapping and measuring their thickness and play an important role in diagnosis of retinal diseases such as age-related macular degeneration (AMD) [43] and diabetic eye disease [44]. OCT can be, further, helpful in diagnosis of glaucoma [45], macular pucker [46], macular edema [47], central serous retinopathy [48], diabetic retinopathy [49], macular hole [50].
Another type of images, discussed by [51], that can be helpful in the diagnosis of age-related macular degeneration (AMD) [52] are fundus images which capture the rear of the eye. It is 2D imaging modality and since glaucoma is a "3D disease", 3D image modality such as OCT is considered more efficient for diagnosis. Fundus images can also be used for detection and grading of hypertensive retinopathy [53]

### 3.4. Magnetic Resonance Imaging (MRI)

Magnetic Resonance Imaging (MRI) modality, described by the National Institute of Health (NIH), US [22], capture 3d anatomical images noninvasively. MRI scanning does not use any radiation which make it an ultimate choice of capturing when frequent imaging is required in the treatment process especially in the brain. MRI is particularly suitable for capturing the soft tissues of the body, but it is more costly as compared to x-rays and CT scanning. MRI can be used to capture different parts of the body for example MRI are used for diagnosis of aneurysms and tumors [54] as well for differentiating between white matter and grey, in brain. MRI can further be used for spinal cord [55] and nerves[56], muscles [57], and ligaments [58].
There is a specialized MRI called functional Magnetic Resonance Imaging (fMRI), which is used for observing brain structure and locating the areas of the brain which are activated during cognitive tasks.[59]

### 3.5. Ultrasound

Radiologyinfo.org for patients [60] described ultrasound as an imaging modality that invasively create image of organs, tissues, and other structures inside the body invasively by using sound waves without using any radiation. Ultrasound can be used to internal organs within the body, noninvasively. For example, capturing the heart, eyes, brain, thyroid, blood vessels, breast, abdominal organs, skin, and muscles. Ultrasound images are captured in 2D, 3D, but it can also capture 4D images which is 3D in motion such as a heart beating [61] or blood flowing through blood vessels [62].

### 3.6. Histopathology or Whole-Slide Imaging (WSI)

The Whole-Slide Imaging (WSI) refers to capturing the microscopic tissue specimens from a glass slide of biopsy or surgical specimen which results in high-resolution digitized images. These images are captured through, first taking small high-resolution image tiles or strips and then montaging them to create a full image of a histological section [63]. Specimens on glass slides transformed into high-resolution digital files can be efficiently stored, accessed, analyzed, and shared with scientists from across the web using slide management technologies. Moreover, WSI is changing the workflows of many laboratories. It is used in various disease diagnostics, prognostic and treatments such as survival prediction [64],detection of tissue phenotypes [65], Automated grade classification [66, 67, 68], segmentation of microvessels and nerves [69, 70], Multi-Organ Nuclei Segmentation [71].

### 3.7. Positron Emission Tomography – Computed Tomography (PET-CT) Scans

According to Radiologyinfo.org [60], Positron Emission Tomography, also called PET imaging or a PET scan, small amounts of radioactive material called radiotracers for capturing images. PET-CT scans can be used for cancer detection and diagnosis [72], determining spread of the cancer, determining the recurrence of cancer, metastasis [73], evaluating brain abnormalities like tumor [36] and memory disorder [74], mapping normal human brain and heart function.

## 4. Deep Neural Networks - Enhancing Representation Learning from CNNs to Vision Transformers

The goal of this section is to bridge the gap between AI and healthcare analysts. It introduces the deep learning concepts, techniques, and architectures that is found in the papers surveyed for this review article.

The progression of deep neural networks in computer vision has contributed to various fields of study, and it primarily revolves around convolutional neural networks (CNN). For instance, while assessing medical images, practitioners can recognize if there is an anomaly. Similarly, this mechanism can be taught to a computer via CNNs to diagnose a disease or an anomaly while taking images as input, hence, giving vision to a computer.The model of a basic CNN is illustrated in Figure 10. It takes the image input as a matrix of pixel values, assigns weights to learn the various differentiable features [15]. It then passes the image through multiple layers and uses multiple filters to capture the discriminated features from the image. CNNs generally consist of three kinds of layers: convolution layers, pooling layers, and full-connected layers [75]. Convolution layers are responsible for learning features and capturing the Spatial and Temporal dependencies between the features by application of relevant filters. The pooling layer is responsible for reducing the size of feature maps to capture more semantic information than spatial information. In convolutional layer filters of size NxN where N is equal to 1,3, 5, 7, or any other odd number. The pooling layer uses a window of size 2x2, 3x3, or any other desired size to take average or maximum value in that window. Before the fully connected layer, the output of the convolutional and pooling layer which is called feature map is flattened to make a fully connected layer at a function such as softmax is applied to make a prediction and a loss function is used to calculate the error and the is back propagated to update the values of learnable parameters.

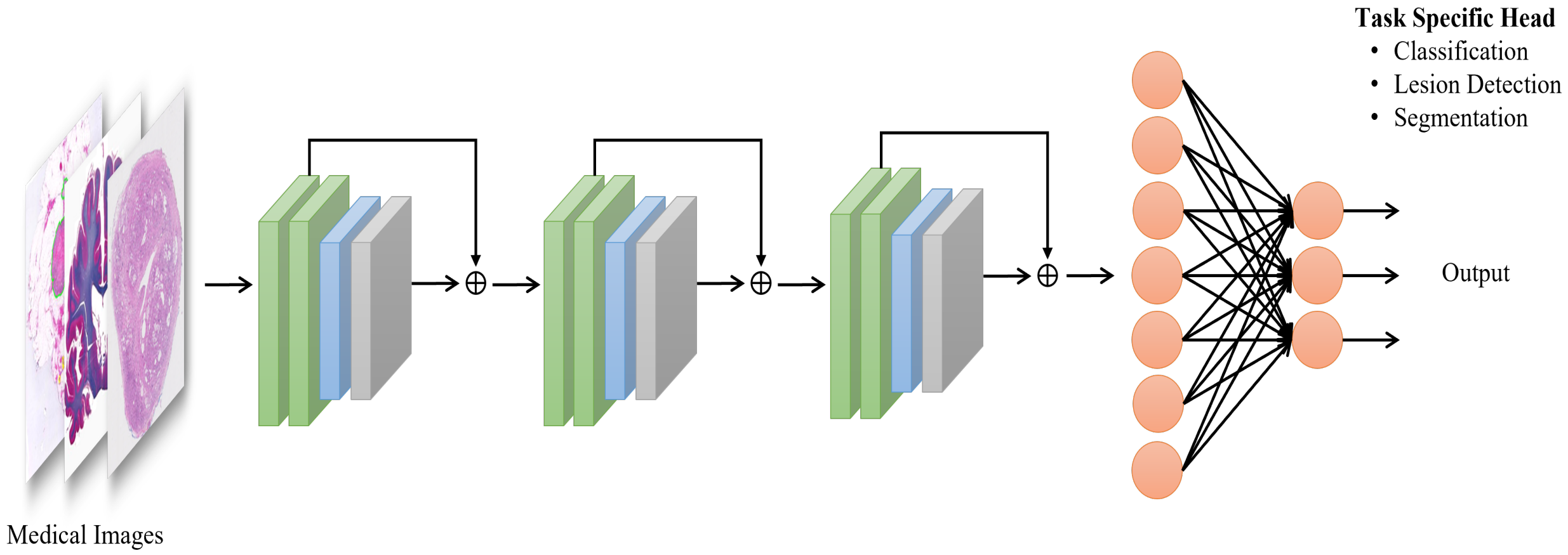

**Figure 10:** A general framework of Convolutional Neural Networks (CNNs)

Depending on the application for example image classification, fully connected layers are added at the end of the network. Stacking these layers on top of each other with a specific arrangement with the help of a differentiable function is known as CNN architecture. In recent years several CNN architectures are developed with various such arrangements: AlexNet [76], VGGNet [77], GoogleNet [78], ResNet [79], ResNeXt [80], Squeeze and Excitation Net [81], DenseNet [82], and EfficientNet [83].

Convolutional neural networks are used in various applications in the categories of image classification, detection, and segmentation, etc. For example face detection [84], identification of emotions [85], Speech recognition, and Machine translation using CNNs [86], etc. Considering the applications of CNN, it can be inferred that they can enable commendable results [87]. They are known to be a black box, as the training is according to the task and domain. One major limitation is the unclarity of results i.e. the reason for a particular outcome. Especially in the medical domain, it is imperative to know the cause of a specific outcome, otherwise, a wrong diagnosis can be a threat to human lives.

One way to tackle this problem head-on is to have such a model that focuses on relevant parts of the image and can be visualized by the doctors. To elucidate this issue, *Attention models* were proposed [16, 69]. The attention model focuses on the parts that are relevant to the input sequence. Moreover, a model was proposed known as Transformers, it used the concept of Attention to enhance the training speed [19]. Transformers consist of multiple blocks of identical encoders and decoders, which were composed of self-attention block and feed-forward networks. In addition, the decoder consists of an extra attention block, which focuses on the relevant part of the sequence. The embedded words of the input were passed to the encoder sequentially and were propagated to all the encoders. The outcomes of the last encoder were then fed to the decoders. The performance of the transfer models was state-of-the-art in the tasks related to natural language processing.

Inspired by the transformer model, Dosovitskiy et al [19], applied it to the images and it can be used to replace CNNs . This model was called Vision Transformers (ViT) and its structure is illustrated in Figure 11. ViT model introduced global attention, but not on the entire image, rather they divided the entire image into small image patches of 16x16. They introduced simple numbers 1, 2, up to n as positional embeddings for specifying the positions of the patches.

A lookup table was introduced which contained a learnable vector against each number representing the position of the patch. These embeddings were passed to the network along with the patch. Each patch was unrolled to a linear vector and projected linearly on embedding matrix. The final embedding along with positional embedding was then fed into the transformer encoder. Along with embedding for patches, an extra embedding with number 0 is also fed into the network and its output was obtained. Thus, Vision transformers have the capability of modeling global context which assists in more accurate results. Lastly, in this review, medical images are considered as the input for vision transformers.

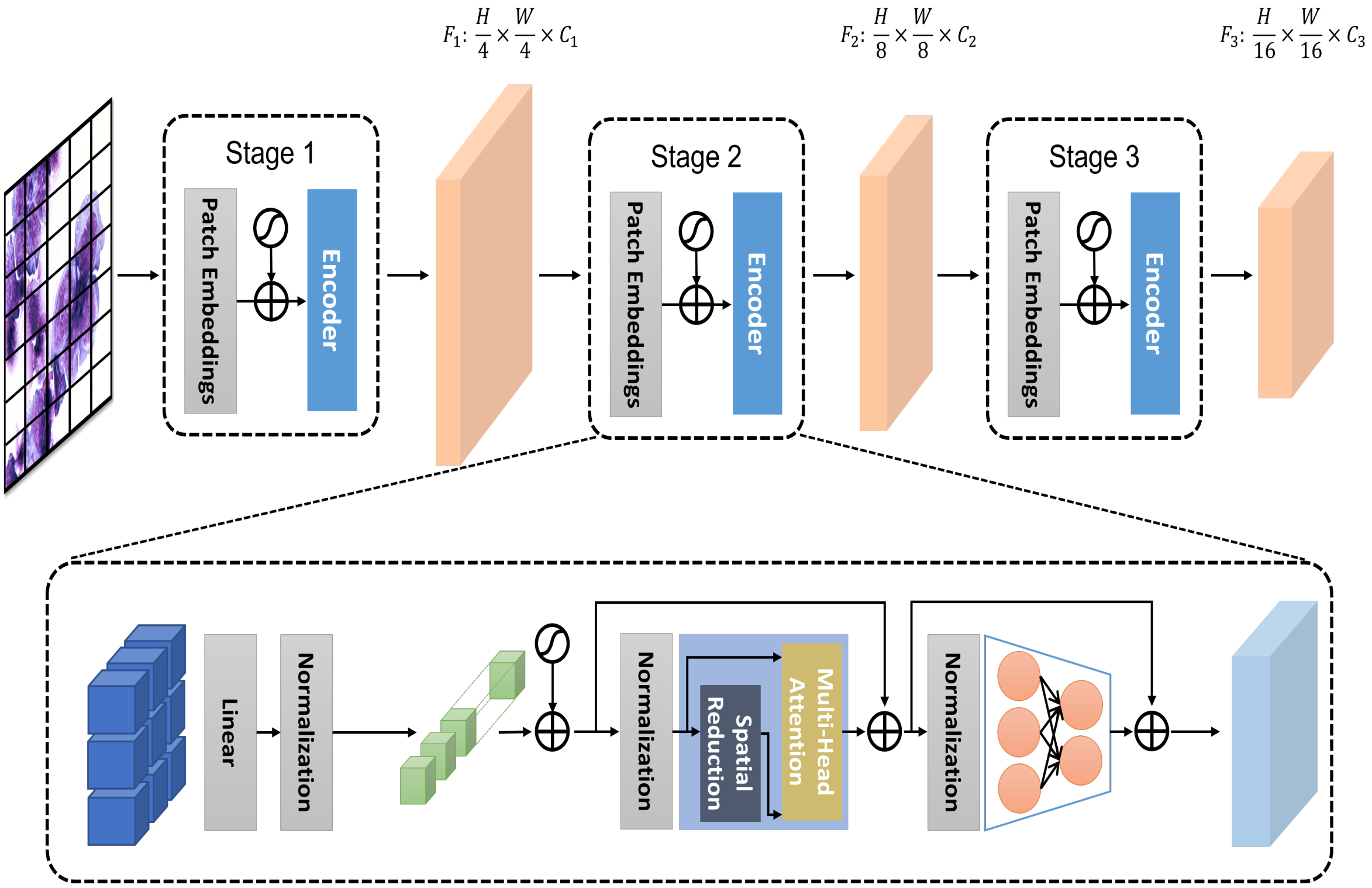

**Figure 11:** Structural representation of Vision Transformers

### 4.1. Open Access Medical Imaging Datasets employed in ViT Applications

In table 3, we have summarized and structured open-accessed datasets in tabular form. The table includes information regarding the tasks i.e. classification, segmentation, detection, report generations, and miscellaneous. Furthermore, the respective image modalities and their applications are also included. Next, we have also compiled the description and links to the corresponding datasets. Hence, it will assist researchers to identify these resources that can be utilized against different applications.

Table 3: Publicly available datasets of medical imaging modalities used by researchers for assisted diagnosis

| Tasks | Modality | Applications | Datasets | Description | Download Links |
|---|---|---|---|---|---|
| Classification | CT Scans | Pulmonary Nodule Characterization | LUNA16 [88] | A commonly used dataset for lung nodule identification and false positive reduction. | Download |
| | | | LIDC-IDRI [89] | One of the largest publicly available lung cancer screening datasets, made up of diagnostic and screening thoracic computed tomography (CT) images. | Download |
| | | Emphysema Classification | Computed Tomography Emphysema Database [90] | A freely available dataset that includes 115 high-resolution CT (HRCT) scans and 168 square patches that were manually annotated in a subset of the slices. | Download |
| | | COVID-19 Detection | COVID-CT [91] | A freely accessible collection of CT-scan images extracted from a number of scholarly articles. | Download |
| | | | Sars-CoV-2 [92] | A multi-class CT scan dataset for identification of SARS-CoV-2 infection. | Download |
| | | | COVD19-CT-DB [93] | COVID19-CT-Database consists of chest CT scans that are annotated for the existence of COVID-19. | Download |
| | | | COVID-CTset [94] | One of the largest open access COVID-19 lung CT dataset that is available online. | Download |
| | X-rays | Pneumonia Classification | Chest X-ray Images [95] | A Dataset of validated OCT and Chest X-Ray images. | Download |
| Classification | X-rays | Tuberculosis Prognosis and Detection | Montgomery County (MC) CXR | Small tuberculosis dataset from USA. | Download |
| | | Tuberculosis Prognosis and Detection | Shenzhen dataset | Small tuberculosis dataset from Shenzhen (China). | Download |
| | | | COVID Chest X-ray dataset | An open access dataset of chest X-ray and CT images of patients which are positive or suspected of COVID-19 or other viral and bacterial Pneumonias. | Download |
| | | Interpretable COVID-19 Detection & Severity Quantification | BIMCV COVID-19+ [96] | BIMCV-COVID19+ dataset is a large dataset with chest X-ray images CXR (CR, DX) and computed tomography (CT) imaging of COVID-19 patients along with their radiographic findings. | Download |
| | | | COVID-19 Posterior-Anterior Chest Radiography Images [97] | This is a curated COVID-19 Chest X-ray image dataset that was created by combining 15 publicly accessible datasets. | Download |
| | | | Extensive COVID-19 X-Ray and CT Chest Images [98] | In this COVID-19 dataset, both Non-COVID and COVID cases are included of both X-ray and CT images. | Download |
| | | | COVIDx Dataset [99] | A database of chest X-ray images for COVID-19 positive cases along with Normal and Viral Pneumonia images. | Download |
| | MRI Scans | Multi-Modal Medical Image Classification | MRNet Dataset [92] | The MRNet dataset consists of 1,370 knee MRI exams performed at Stanford University Medical Center. | Download |
| | Fundus Images | Retinal Image Synthesis and Disease Prediction | Color Fundus Images [100] | A database that has both FFA image and color fundus image in this DR grading system. | Download |
| | Histopathology Images | Colorectal Histopathology Image Classification | Colorectal Cancer Histology Dataset [101] | This data set represents a collection of textures in histological images of human colorectal cancer. | Download |
| Detection | X-rays | COVID-19 Diagnosis | COVIDx [99] | A database of chest X-ray images for COVID-19 positive cases along with Normal and Viral Pneumonia images. | Download |
| | | | COVIDGR-E [102] | It is built by adding 426 pneumonia images from the ChestX-ray8 database to the COVIDGR-1.0 dataset | Download |
| | Fundus Images | Microaneurysms Detection | IDRiD [103] | This dataset consists of 81 images. | Download |
| | CT Scans | COVID-19 Diagnosis | COV19-CT-DB [93] | COVID19-CT-Database consists of chest CT scans that are annotated for the existence of COVID-19. | Download |
| | | | COVIDx-CT-2A [104] | A benchmark dataset: the largest comprising a multinational cohort of 4,501 patients from at least 15 countries and contains three classes – COVID-19 Pneumonia, non-COVID-19 Pneumonia, and normal. | Download |
| | Histopathology Images | Cancer Detection | The Cancer Genome Atlas [105] | The compendium includes heat maps for 33 different tumor types and three platforms: gene expression, reverse-phase protein arrays (RPPA), and miRNA expression. | Download |
| Segmentation | | | iSeg-2017 [106] | this datasets contains brain MRI's of 39 subjects | Download |
| Segmentation | MRI Scans | Brain Tissue Segmentation | BraTS-2020 [107] | It contains 1756 MRI of 439 subjects | Download |
| | | | MRBrainS [108] | It contains 20 Fully annotated multi sequence 3T MRI Brains | Download |
| | | | UKBB [109] | | Download |
| | MRI Scans | Cardiac Segmentation | M&MS-2 | It contains 360 subjects having 200 training and 160 testing images | Download |
| | | | ERI [110] | It contains LGE Data of 28 patients. the number of segmented images are 358 | Download |
| | | Abdominal Segmentation | CHAOS [111] | This Data set contains images for T2 Segmentation of Liver and Kidneys | Download |
| | X-rays | Tooth Root Segmentation | DRIVE [112] | It include 40 color fundus retinal images that are randomly selected | Download |
| | | Knee Segmentation | OAI | 212 Knee images were segmented randomly and the training and testing split was 100 and 112 | Download |
| | | Lung Segmentation | JSRT [113] | The database includes 154 conventional chest radiographs with a lung nodule (100 malignant and 54 benign nodules) | Download |
| | CT Scans | Pediatric Segmentation | KiTS19 [114] | This Dataset contains renal Tumor Segmentation Images of 210 Adults | Download |
| | | Drusen Segmentation from Retinal OCT Images | USCD [115] | This dataset contains 8616 retinal OCT B-scans | Download |
| Clinical Report Generation | X-rays | | IU Chest X-ray Collection [116] | A public and open source OpenI or Indiana University Chest X-rays database which contains 3955 medical reports with 7470 frontal and lateral chest x-ray images. | Download |
| | | | MIMIC-CXR [117] | It is the recently released largest dataset, consists of 377,110 chest X-rays images and 227,835 reports from 64,588 patients. | Download |
| | | | PEIR GROSS [118] | It consists of publicly accessible 7442 teaching images, spread across 21 predefined subcategories. The vocabulary size of the total image captions is 4,452. Each image on average contains a 12 word caption. | Download |
| Miscellaneous | PET-CT Scans | Medical Image Enhancement | NIH-AAPM-Mayo Clinic LDCT Grand Challenge [119] | A public and open source 30 contrast-enhanced abdominal CT patient scans. | Download |
| | | | Kirby21 Dataset (KKI01 to KKI05) [120] | A public and open source dataset containing correlation data for 20 subjects from Kennedy Krieger | Download |
| | MRI Images | Medical Image Reconstruction | DIVerse 2K resolution high quality (DIV2K) images dataset [121] | It contains a total of 1000 2K resolution RGB images. | Download |
| | | | fastMRI Scans [122] | T1- and T2-weighted images from 150 subjects were analyzed (100 for training, 10 for validation, 40 for testing). | Download |
| | Fluorescence Microscopy | Denoising of Celullar Microscopy Images for Assisted Augmented Microscopy | Flywing, Planaria and Tribolium datasets [123] | The training data 17,005 and 14,725 small cropped patches of size 64×64×16 for Planaria and Tribolium datasets, while the testing data are 20 testing images of size 1024×1024×95 and 6 testing images of average size around 700×700×45 for the two datasets, respectively | Download |

# 5. Visual Recognition Tasks in Medical Images

Artificial intelligence and deep learning have played a vital role in assisting clinicians for better diagnosis. In this context, the application of CNNs and ViTs on medical images assists the healthcare professional in disease classification, lesion detection, segmenting the anatomical structures, automated report generation, denoising the images, medical image registration, and various other tasks. This section gives a brief overview of the above-mentioned visual recognition tasks performed by the application of CNNs and ViTs on medical images.

## 5.1. Medical Image Classification

The practitioners give their diagnosis by analyzing the medical images, hence, determining the presence and type of the disease. This conventional diagnosing way can be assisted with deep learning techniques. Figure 12 shows a generalized classification network. Through these techniques, the ambiguity in diagnosis among different doctors can be mitigated and the outcomes will be more accurate. Thus, results achieved through CNNs are not only time efficient but can also assist healthcare professionals. Using CNN models, an input image is fed into the network, which is then assessed, analyzed, and interpreted to determine the target and object of different modes [124]. This process is known as image classification. For example, considering the figure 5, the chest X-ray images are given as input, and the model classifies them into normal and pneumonia images. At present, image classification has various applications in the medical domain such as; skin cancer [125], diabetic retinopathy [126], tuberculosis [127], etc.

The significance of image classification using CNNs can be determined by the aforementioned applications. These applications were achieved through various CNN architectures such as AlexNet [76], VGGNet [77], GoogleNet [78], ResNet [79]. Later, more resource-efficient architectures were proposed i.e. MobileNet [128] , Squeeze and Excitation Net [81], and EfficientNet [83], etc. Through these Convolutional neural networks, commendable results were achieved in medical applications, however, in terms of resources, improvements are still required.

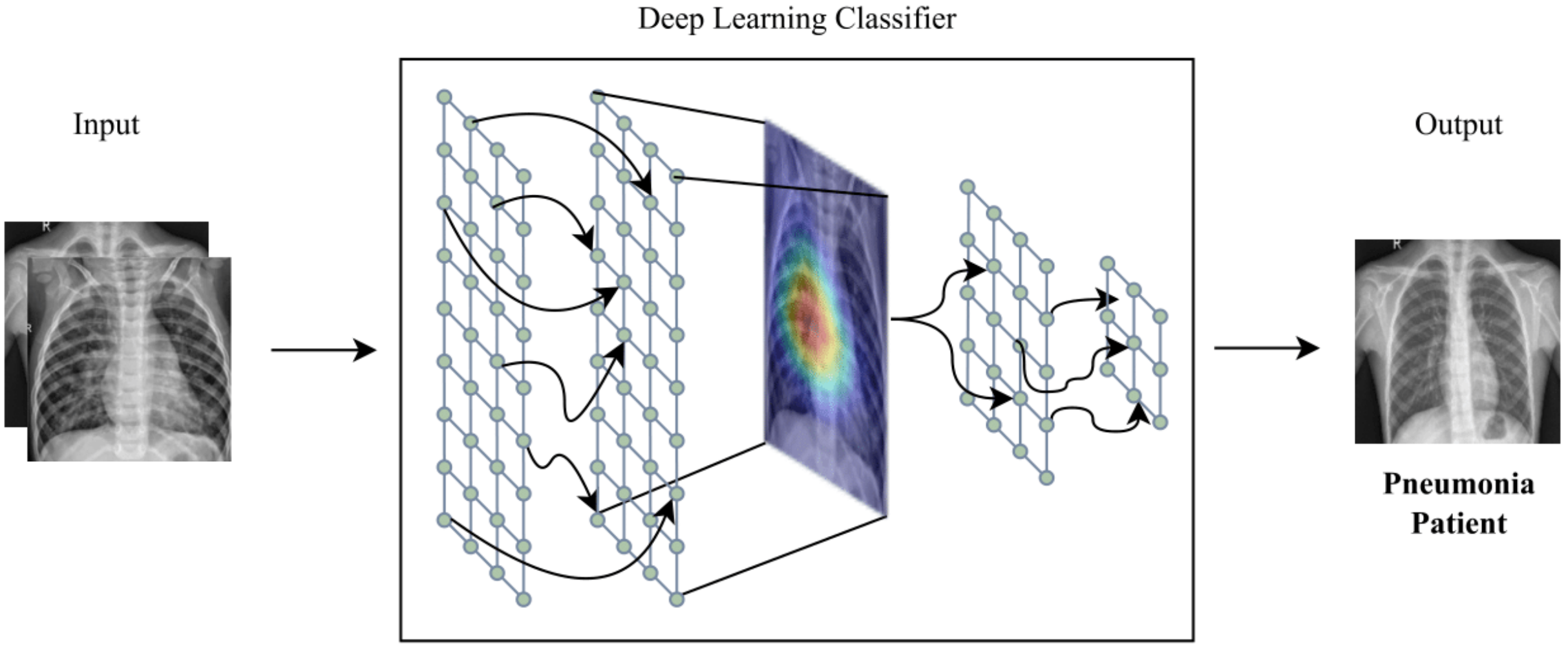

**Figure 12:** Medical image classification pipeline in which Chest X-rays (CXRs) are fed into a Deep CNN architecture, which assigns each image a class, such as pneumonia patient or healthy patient.

## 5.2. Lesions Detection

Classifying images was indeed a step forward towards automated diagnosis. Nevertheless, locating the object plays an important role while developing a more functioning application. In the field of computer vision, one of the underlying goals is to classify the object into a category and determine the location of the object in a given image, this technique is known as object detection. Figure 13 figuratively explains the a general detection network. For instance, if the input image is an X-ray of a hand, the object detection model will not only classify the category i.e. fractured bone but also localize it by using bounding boxes. Considering a situation, where bones were fractured on multiple locations, here, object detection will be a better technique to opt for as it will be more helpful to the diagnostician. The applications of object detection involve; face detection [84], plant disease identification [129],weapon detection [130] , emotion detection [85], etc.

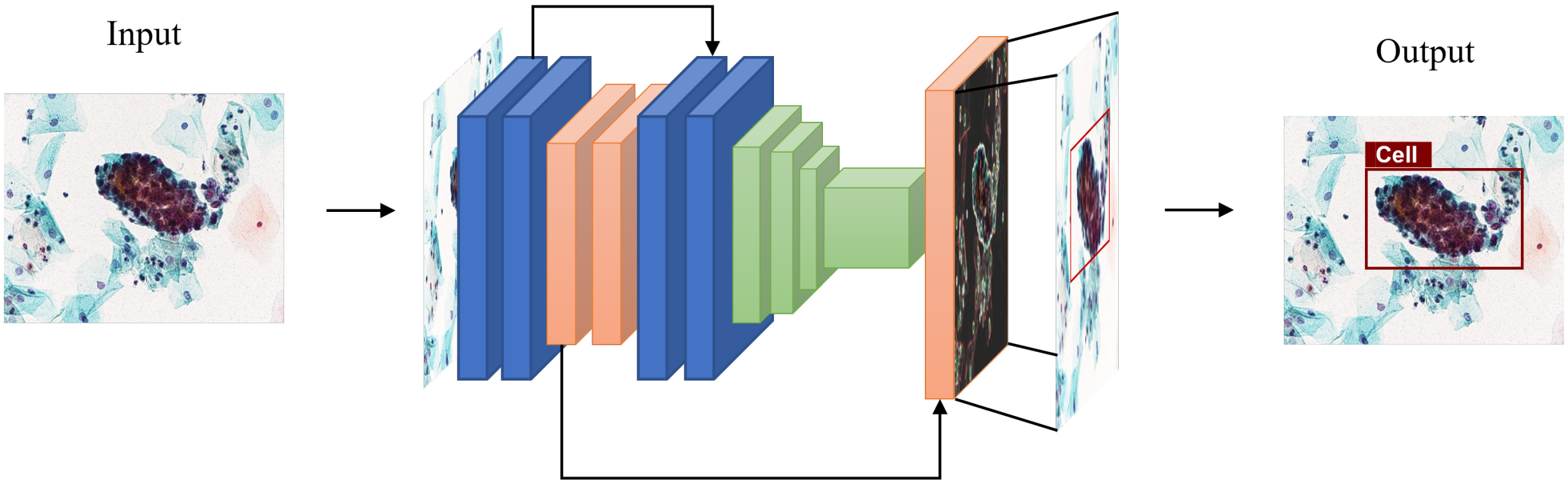

**Figure 13:** A CNN architecture detecting a colony of tumorous cells given a histopathology image.

The task of object detection is composed of two types; two-stage networks and one-stage networks [124]. The two-stage networks are based on region proposal algorithms such as R-CNN [131], Fast R-CNN [132] , and Faster-RCNN [133]. The other technique is designed in a way that it works directly on images, examples include; YOLO [134], SSD [135], etc. The two design types have a trade-off between accuracy and time efficiency. The two-stage networks are capable of more accuracy, whilst, one-stage networks have more speed. Thus, it depends on the task at hand and dataset, while choosing these networks for object detection. The limited datasets in the medical domain are the biggest constraint while training the models, therefore, researchers are working on models which can work with limited datasets.

### 5.3. Anatomical Structure Segmentation

In the medical domain, there are numerous cases where it is difficult to distinguish between two different lesions as there are minor differences. Since lesions can have different treatment strategies, determining them as separates is vital. It is indeed a challenge to recognize such subtle differences, but it is not impossible. If the images are classified pixel-wise, they can be localized as well, and result in a fine outline of the object. Such a technique is known as Image segmentation. For example, if an MRI image is fed into the model, the outcome will not just classify the tumor type, its anatomical structure will also be highlighted as seen in Figure 14. There are various applications of image segmentation, which includes ; Cardiovascular structure [136] , prostate cancer [137], blood vessel [138] etc.

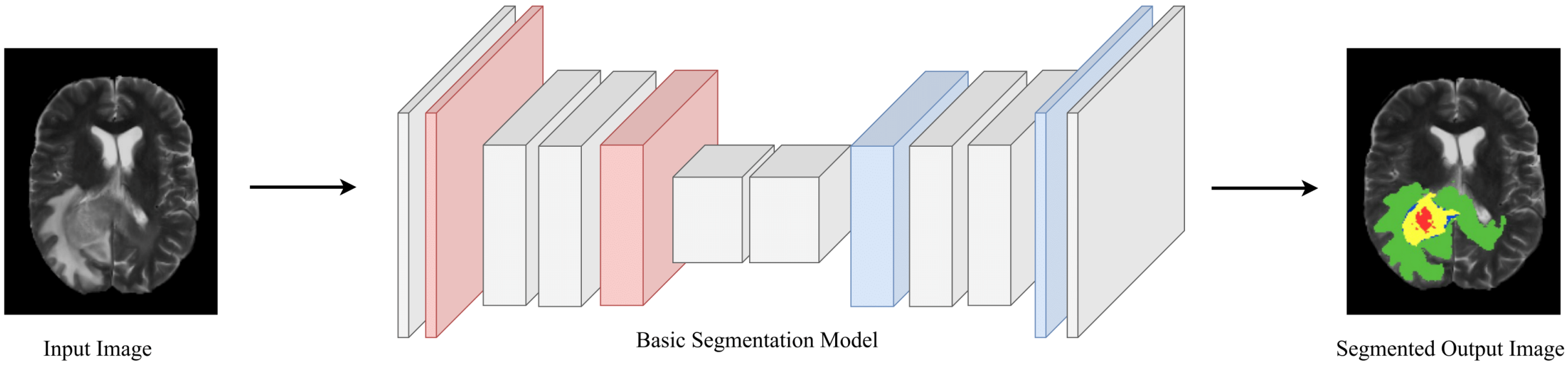

**Figure 14:** A segmentation framework in which brain MRI scans are fed into a deep CNN architecture which is not just classifying and locating the tumorous region but also highlighting the anatomical structure.

Fine-grained segmentation is a decisive step in image-guided treatment and computer-aided diagnosis. The widely used architectures for image segmentation are; U-Net [139], DeepLab [140], Masked R-CNN [141] etc. The usage of these segmentation models depends on the problem that is being solved. For instance, multi-scale objects in the image, deep lab, and its various structures will be a wise choice. Lastly, the concern regarding image segmentation is the lack of labeled data due to which researchers are considering more unsupervised approaches, however, it is still a work in progress [142].

### 5.4. Clinical Report Generation

In the healthcare domain, while examining radiology images i.e. Chest X-rays, CT Scans, MRI, etc, the doctors have to write detailed reports of the assessment. This conventional method of report writing is tedious, monotonous, time-consuming, and error-prone for radiologists [143]. Immense progress has been made in deep learning to generate medical reports automatically. Automatic report generation can assist clinical practitioners in quick and accurate decision-making [144]. For example; if the CT scan of a brain is given as input, the output would be a complete report such as, if the tumour exists, the location of the tumour, its size and other details. Figure 15 shows a general workflow of clinical report generation. Medical report generation is an application of image caption in which these models are applied to medical data. Image captioning refers to computers generating captions by giving images as input. It consists of mainly an encoder-decoder where CNN is used to extract features, and LSTM or RNN are used to generate the captions [145]. The initial work on its architectures include Show and Tell [146], Show Attend and Tell [147], Neural Talk [148], etc. Some of the State of the Art include DenseCap [149], SemStyle [150], Dense-CaptionNet [151], etc. Lastly, as in supervised learning, a large amount of annotated data is required for these models to perform well, in the future unsupervised learning could be a more powerful way to proceed with them [152].

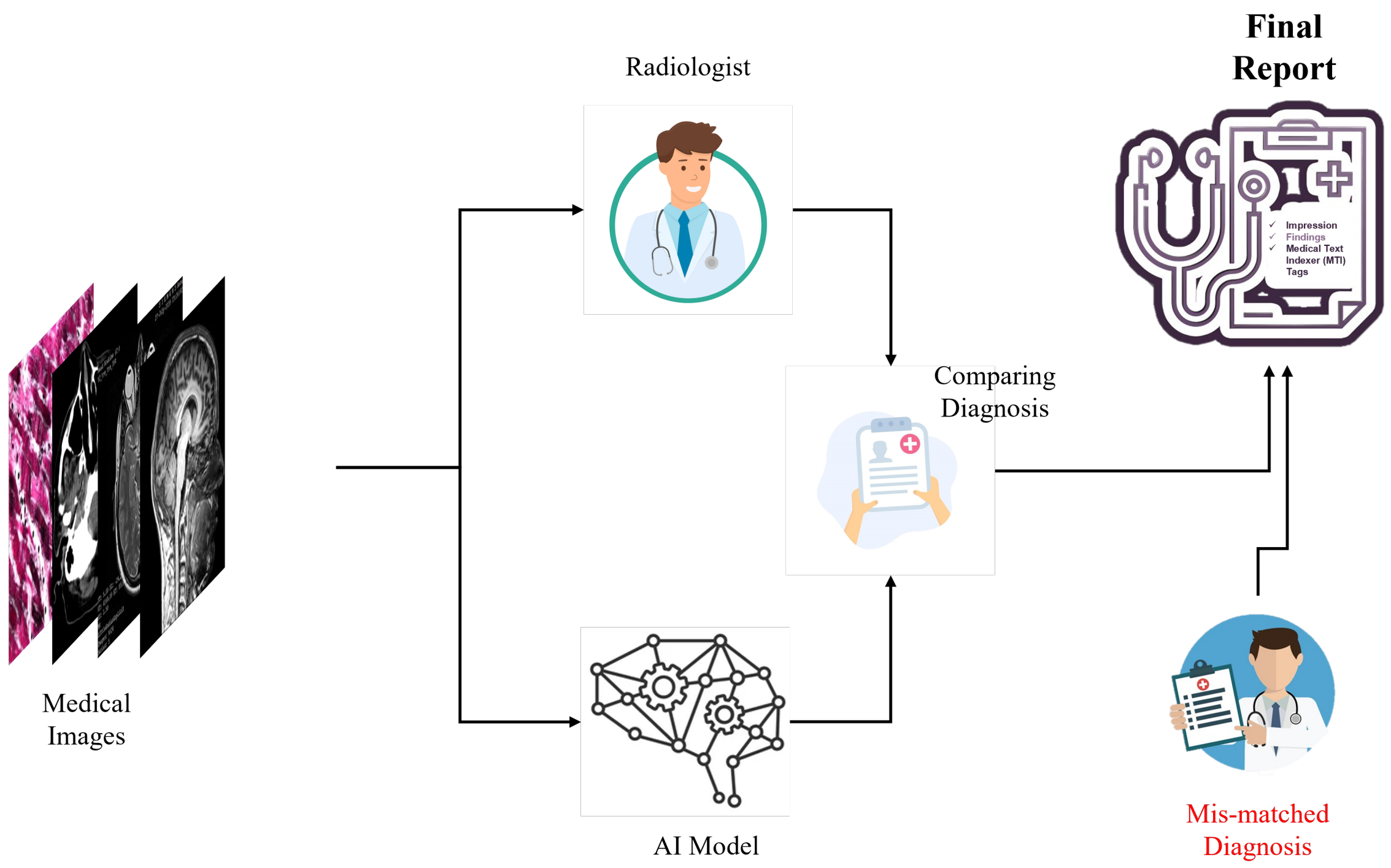

**Figure 15:** A visual representation of clinical report generation mechanism in which different types of image modalities are independently diagnosed by Radiologist and AI model, and then compared to each other to provide a precise medical report.

# 6. A Recapitulation on Vision Transformer Application to Medical Image Analytics for Assisted Diagnosis

## 6.1. ViT in Image-based Disease Classification

Deep learning has recently come to the power in a variety of research domains. Convolutional Neural Networks (CNNs) have been the most dominant deep neural networks for autonomous medical image analysis applications such as image classification during the last decade. These models, however, have shown poor performance in learning the long-range information, due to their localized receptive field, which limits their capabilities for vision related tasks.
Transformer architecture, proposed by Vaswani et al. [16], is currently the most popular model in the field of natural language processing (NLP). Getting inspiration from the success of self-attention based deep neural architectures, Dosovitskiy et al. [19] introduced Vision Transformer (ViT) model for image classification based applications. In these models, the overall training process is predicated on dividing the input image into patches and considering each embedded patch as a word in NLP. Self-attention modules are used in these models to learn the relationship between the embedded patches.
In the following section, we will take a step forward in exploiting the potential applications of self-attention based architectures like Vision Transformers (ViT) for the task of medical image classification such as COVID-19 detection and severity quantification, Emphysema classification, tuberculosis prognosis etc. The section is further divided into medical imaging modalities, with a focus on contributions made by vision transformers to the respective medical applications. The section ends with a tabular summary of the proposed approaches and their performance matrices.

### *6.1.1. Computed Tomography (CT) Scans:*

*Pulmonary Nodule Characterization:* Lung cancer is one of the most frequently reported causes of cancer-related morbidities and mortalities [153]. Early detection and treatment of multiple pulmonary nodules has become a challenge in clinical practice as it is one of the most efficacious ways to reduce the number of fatalities associated with the condition. Prior research [154, 155, 156] on detection and characterization of lung nodules focused on learning the associations between various nodules. In other words, they particularly use solitary nodule approaches on several nodular patients, ignoring the relational/contextual information. To overcome this issue, Yang et al. [157] proposed a Multiple Instance Learning (MIL) strategy, which empirically proved the utility of learning the relationships between nodules. It's the first time researchers have looked into the relationships between several lung nodules and extracted critical relational information between solitary-nodule voxels. Instead of using typical pooling-based aggregation in multiple instance learning, they created Set Attention Transformers (SATs) based on self-attention to understand the relationships between nodules. A 3D DenseNet is employed as the backbone to learn representations of voxels at the solitary-nodule level. The SATs are then used to determine how several nodules from the same patient are related. This data-driven methodology might aid in understanding of etiologic and biologic processes as well as metastasis diagnosis of multiple pulmonary nodules and motivate clinical and biological research on this important topic.

*Emphysema Classification:* Chronic obstructive pulmonary disease (COPD) is a heterogeneous disorder with a variety of characteristics, including small and large respiratory inflammation, as well as Emphysema, which is the most common cause of progressive lung tissue loss. Emphysema, as characterized by the destruction and persistent growth of the alveoli, can be classified automatically which can aid in determining and quantifying lung destruction patterns. In this regard, Convolutional Neural Networks (CNNs) serve an essential role, particularly in pulmonary CT image classification, but transformers have yet to be explored. As a result, Wu et al. [158] conducted a thorough assessment and extensive evaluation of the ViT model for Emphysema classification. First, large image patches (16 x 16) are cropped from CT scans. After resizing, the patches are flattened and linearly embedded to create a sequence of patch embeddings. The positional information is kept by concatenating the class embeddings with the patch embeddings. To acquire the representation, the final embedding sequence is passed into the transformer encoder module. Finally, the learnable class embedding is fed to a softmax layer for Emphysema classification. Despite the fact that this study employed a vision transformer model to classify emphysema, unlike other techniques that use CNN models, it still has several limitations. For example, patch-based classification may not be as convenient as pixel-based segmentation. Furthermore, the architecture just uses the transformer encoder block, and the CNN's benefit is not utilized. In short, Emphysema quantification is difficult, and classification is merely the first step. Proposing more efficient networks capable of learning semantic information of Emphysema by partial or accurate annotations may be a pressing need. In the near future, more research on the segmentation and quantification of Emphysema will be conducted.

*COVID-19 Detection:* The infectious Coronavirus (COVID-19) and lung disorders have been at the vanguard of the research community as the pandemic has caused significant public health concerns throughout the world. Using computer vision methods, several attempts [159, 160, 161] are being undertaken to create automated systems for faster and more effective diagnosis of COVID-19. As per several research studies [162, 163], some radiographic manifestations, such as broncho vascular thickening, Ground Glass Opacities (GGO), crazy-paving pattern, and consolidation, have been found in chest CT images. However, with the rapidly growing number of patients in the current situation, radiologists have a significant challenge in manually interpreting CT scans.

Ambita et al. [164] were the first to use a vision transformer to the task of COVID-19 detection from computed tomography (CT) scans. They implement a variety of vision transformers (e.g., ViT-B 16, ViT-B 32, ViT-L 16, ViT-L 32, and ViT-H 14) for image classification. They employed ResNet-based Self-Attention Generative Adversarial Network (SAGAN-ResNet) as a data augmentation approach for synthetic image generation to alleviate the problem of insufficient data. Furthermore, they demonstrated how ViT offers visualizations for the images by exhibiting which sections of the input image the model focuses its attention on in the various layers. This might be useful for radiologists when analyzing CT scans. Improvements may also be made by evaluating the proposed approach on different datasets and customizing the architecture of transformers or GAN.

Zhang et al. [165] attempted to broaden the scope of vision transformers such that they may be used as a robust feature learner for COVID-19 diagnosis in 3D CTs. Inspired by the success of Swin vision transformer and CT classification work in [166, 167], their framework is made of two key stages: lung segmentation followed by image classification using Swin transformer as a backbone. In the first stage, a pre-trained Unet is used for lung segmentation in CT scans and produced a lung mask that restricted learning to certain lung regions. The features from each 2D CT slice are then extracted using a Swin vision transformer, which are subsequently aggregated into 3D volume level features using a max-pooling layer. However, it's worth noting that the framework equipped with the backbone of EfficientNetV2-M achieves a good speed-accuracy tradeoff according to the results on the validation dataset. This implies that in future study, merely increasing the model size might result in an improvement in classification.

Above reviewed studies are evident that vision transformers (ViT) is a novel and quickly evolving approach that has demonstrated excellent results using COVID-19 datasets. Than et al. [168] conducted a preliminary study to investigate the efficacy of using ViT with different sized patches on CT scans of diseased lungs, COVID-19 infected lungs, and normal lungs. A default positional embedding is used and the combination is then passed to a transformer encoder which is composed of alternating layers of multi-headed self-attention and multi-layer perceptron (MLP) units. The transformer encoder's output is sent into an MLP head, which outputs the predicted class. The proposed methodology is simpler and less demanding on computational resources as compared to CNNs, however, pretraining the transformer and adding a convolutional layer in front of it (i.e. Convolutional vision transformers) may increase performance. Aside from that, other hyper-parameters can be modified to increase performance, and explainable artificial intelligence (XAI) will be used in the future to explain how deep learning networks like ViT make decisions.

COVID-19 CT scans contain not only the local features, such as local crazy-paving and local hemangiectasis, but also have global characteristics. Since it is characterized by the combination of both local and global features, extracting image features with relatively complex features is a challenging classification problem of such medical imaging modalities. Fan et al. [169] proposed a parallel bi-branch network (TransCNN Net) that is essentially based on the Transformer and Convolutional Neural Network. Unlike the conventional approaches, the size of the convolutional filter kernel is not changed to extract features at different scales; instead, they use the global receptive field of the transformer network. A bi-directional feature fusion structure is then designed, which fuses the extracted features from two branches bi-directionally, forming a network that has the potential to extract more comprehensive and ample features.

#### 6.1.2. X-ray or Radiographic Images:

*Pneumonia Classification:* Pneumonia is an infectious disease in which the alveoli in the lungs is to be filled with fluid or pus, making it painful to breath as well as decreasing oxygen intake. A detailed inspection of chest X-ray images by the radiographer or radiotherapist is required to diagnose Pneumonia. As a result, pneumonia detection is a time-consuming process, and even a slight error can result in an excruciatingly painful outcome. Several researchers have explored various computer vision approaches to diagnose Pneumonia, using X-Ray images of human chests. Tuagi et al [170], have proposed a vision transformer (ViT) model for the classification of pneumonia. Further, to validate the performance of the proposed model, it was also compared with the CNN model and VGG16. It was observed that ViT outperformed other techniques reaching the highest accuracy of 0.96. In addition, it outperformed other models in terms of computational cost as well. It is observed that the model still requires further experimentations to investigate the robustness of the model with more heterogeneous data. There hasn't been a lot of work done using vision transformers in the field of chest x-ray diagnosis. It may be useful in the diagnosis of other disorders, such as Covid-19 detection, Cystic Fibrosis or Emphysema, Edema, Pleural thickening, Effusion, and even Cancer.

*Tuberculosis Prognosis and Detection:* Tuberculosis refers to an infection that affects the lungs and can travel to other parts of the body. It can be diagnosed and assessed by referring to chest x-rays. If the infection is cured at an earlier stage, it can save a person from further misery of painful treatment. For the classification of infected and non-infected chest x-rays, a methodology was introduced by Duong et al. [171]. The authors have used EfficientNet with vision transformers for the detection of tuberculosis using chest X-rays images. The performance metrics include; accuracy, precision and recall, f1-score, and area under the curve. The highest accuracy achieved was 97 percent with the backbone of efficient net B1. Hence, the paper validates the robustness of vision transformer models used on a heterogeneous dataset. However, it should be further evaluated on different baselines models.

*Interpretable COVID-19 Detection & Severity Quantification:* The novel coronavirus disease 2019 (COVID-19), caused by the severe acute respiratory syndrome coronavirus 2 (SARS-CoV-2), has become one of the deadliest viruses of the century as of April 2021, infecting over 137 million people with over 2.9 million death worldwide. In the context of unprecedented COVID-19 pandemic, public health systems have been hit by a slew of challenges, including scarcity of medical resources, that are pushing healthcare workers to face the threat of infection. Deep learning and Computer Vision are commonly employed in numerous fields of medical imaging for the diagnosis of COVID-19 from radio-graphic images, X-rays or CT scans. Even though these techniques have yielded commendable outcomes, the cost is always considered as an important factor for the system to be applicable. The use of a computed tomography (CT) scan for COVID-19 diagnosis offers great sensitivity and specificity [172], but it is a severe burden due to its high cost and risk for cross-contamination in the radiology suite. In comparison of CT Imaging, X-rays have been widely utilized for COVID-19 screening, as they require less imaging time, are less expensive, and X-ray scanners are generally available even in remote regions [173].

Hence, many researchers have worked on diagnosing COVID-19 using Chest X-ray (CXR) images and through these automated methods, counter-check the PCR test results. For instance, in the study presented by Park et al. [174], vision transformers were utilized for both classification and quantification of the severity of COVID-19. Firstly, the low-level features were extracted using state-of-the-art CNN architectures. Secondly, these extracted features were given to the transformer model for classification and severity measurement. Thirdly, the severity was demonstrated through heat maps which gives interpretability to the generated results. The robustness of the model was illustrated through various experimentations on different baseline models and tested on external datasets. The model was evaluated on metrics including; AUC, sensitivity, specificity, and accuracy. The highest accuracy achieved on the external dataset was above 84.8 percent.

Similarly, the study presented by Shome et al. [175] have proposed a pipeline based on Vision Transformers for the classification of COVID-19. The model was also compared with other baseline models to validate the robustness of vision transformers. As compared to [174], it was trained on a larger and more heterogeneous dataset achieving the accuracy of 98 percent and AUC of 99 percent in the binary classification. Furthermore, for the multi-classification which includes pneumonia x-rays as well, the model achieves the accuracy and AUC of 92 percent and 98 percent, respectively. In addition, the model uses Grad-CAM for the visualization through heatmaps, for the explainability of the outcome. Although the model performed better than [174], the model does not quantify the severity of the infection. Further, this research could be expanded to predict the rate at which the infection can spread.

Since the medical data in most domains is inadequate, it can be difficult to build robust models. Considering this issue, Rahhal et al. [176] have put forth a methodology that performs well with small training data. To diagnose covid-19 in both CT scans and X-ray images, the proposed method uses vision transformers as a backbone with the employment of a Siamese encoder. The input image with a corresponding augmented image was fed into the siamese encoder, it was then connected to two classifiers. Further, the output of these classifiers was fed into a fusion layer for the outcome, followed by heatmaps on different layers to interpret the results. Moreover, the siamese encoder caters to the issue regarding deficient data. The proposed model has achieved an accuracy of 94.2 percent which is commendable in limited training data.

Until now, only a few CAD-based approaches for detecting COVID-19 have been developed, but their effectiveness has been hampered due to a number of factors. Taking an inspiration from Convolutional Block Attention Module (CBAM), Nawshad et al. [177] proposed an attention-based Convolutional Module using ResNet32 as the backbone architecture with a 97.69% accuracy. They also conducted a comparative assessment of a variety of deep learning models, including VGG, ResNet, and Xception, for the successful detection of COVID-19 and viral pneumonia. Utilizing the attention module with various CNN-based architectures produced much better results than using the base CNN architectures.

Since it is believed that the newly developing pathogen would have similar low-level CXR features with existing diseases, the unique concept of producing higher-level diagnoses by aggregating low-level feature corpus may be used to swiftly construct a robust algorithm against it.

#### *6.1.3. Magnetic Resonance Imaging (MRI):*

*Multi-Modal Medical Image Classification:* The inadequacy of medical data is discussed in the aforementioned papers, and it can be inferred that researchers are trying to overcome the issue by modeling various architectures. For instance, Dai et al. [178], put forth a hybrid transformer model for the classification of multi-modal images. The pipeline consists of a CNN to extract low-level features, and then transformers are utilized for the global context. The model was applied on two different datasets: for classification of parotid gland tumors and classification of a knee injury. The highest accuracy achieved by the Parotid gland tumor dataset and knee injury dataset was 88.9 percent and 94.9 percent, respectively. Nevertheless, the study has not presented any means for interpretability of the model, as it is essential, especially in the medical field.

#### *6.1.4. OCT or Fundus Images:*

*Retinal Image Synthesis and Disease Prediction:* For the diagnosis of retinal abnormality, Fluorescein Angiography (FA) can be used. To capture the vascular structure of the retina, a fluid is injected into the bloodstream. However, several reactions have been reported with the usage of FA. Another approach to diagnosing abnormality is by using Fundus images. The vascular structure of the eye is not captured in these images. In the paper [179], the authors used fundus images and synthesize them to produce FA's using a generative adversarial network(GAN). Next, these images are then given to a transformer model to classify a normal and abnormal retina. The model achieved the highest accuracy, sensitivity, and specificity of 85.7, 83.3, and, 90.0, respectively. Moreover, this model needs to be validated with a more heterogeneous dataset, along with applying this model for predicting other retinal diseases as well.

#### 6.1.5. *Histopathology Images:*

*Colorectal Histopathology Image Classification:* Colorectal Cancer (CRC) is a type of cancer that begins in the rectum or colon and is defined by the uncontrolled growth of aberrant cells with the latent ability to invade other tissues. Despite the fact that manual inspection of histopathology slides is still crucial in experimental practise, automatic image processing enables the quantitative and rapid analysis of malignant tissues. Early detection is crucial for identifying the appropriate treatment approach and increasing the patient's chances of survival. As a result, automated techniques are needed to save time and eliminate the risk of human error. Artificial intelligence has recently been put to use in the diagnosis and prediction of several forms of cancer. Zeid et al. [180] used Vision Transformers to perform a multiclass tissue classification of colorectal cancer, highlighting the potential of employing Transformers in the histopathological image domain. First of all, a standard vision transformer model was proposed and it achieved an accuracy of 93.3 percent. Since vision transformers demand more data, a hybrid approach combining CNN and transformer was developed. Low-level features are extracted using the CNN model, and the embeddings are sent to the transformer. This model is known as a Compact Convolutional transformer, and it achieved an accuracy of 94 percent. However, experimentation with various datasets and different forms of cancer may also be done to improve the model's overall performance. Deep learning algorithms are now becoming increasingly essential for the identification and classification of colorectal histopathology images. Existing techniques, on the other hand, are more concerned with end-to-end automatic classification using computers than with human-computer interaction. Hence, Chen et al. [181] presented an IL-MCAM framework. It is based on interactive learning and attention techniques. Automatic learning (AL) and interactive learning (IL) are two steps in the proposed IL-MCAM system (IL). To extract multichannel features for classification in the AL stage, a multi-channel attention mechanism model with three separate attention mechanism channels and convolutional neural networks is implemented. The proposed IL-MCAM system constantly adds misclassified images to the training set in an interactive method during the IL stage, improving the MCAM model's classification ability. To handle different colorectal histopathological image classification tasks in the future, permutation and combination can be used to identify the best model for the current task from attention mechanisms and deep learning models. Moreover, attention mechanisms can also be included in various locations of a deep learning model, in order to investigate the influence of convolutional layers on classification performance.

The table 4 given below summarized the performance gain by the reviewed articles of the classification category.

Table 4: List of datasets and performance measures employed by researchers for medical image classification.

| **Modality** | **Publication** | **Dataset** | **Performance Measures** |
|---|---|---|---|
| CT Scans | Yang et al. [157] | LUNA16 [88] | CPM Score = 0.916 |
| | | LIDC-IDRI [89] | Accuracy (%) = 93.17 |
| | Wu et al. [158] | Computed Tomography Emphysema Database [90] | Accuracy (%) = 95.95 |
| | | | Precision (%) = 98.0 |
| | | | Recall (%) = 97.1 |
| | | | Specificity (%) = 98.6 |
| | | COVID-CT [91] | Accuracy (%) = 87.19 |
| CT Scans | | | Positive Precision (%) = 89.11 |
| | | | Sensitivity (%) = 85.71 |
| | | | F1 Score = 0.8738 |
| | | Sars-CoV-2 [92] | Accuracy (%) = 95.41 |
| | | | Positive Precision (%) = 94.30 |
| | | | Sensitivity (%) = 98.03 |
| | | | F1 Score = 0.9613 |
| | Zhang et al. [165] | COV19-CT-DB [93] | Accuracy (%) = 94.3 |
| | | | Precision (%) = 93.7 |
| | | | Recall (%) = 93.8 |
| | | | F1 Score = 0.935 |
| | | | Macro F1 Score = 0.94 |
| | Than et al. [165] | COVID-CTset [94] | Accuracy (%) = 95.36 |
| | | | Senstivity (%) = 83.00 |
| | Li et al.[182] | Private dataset from eight different Hospital [182] | Macro F1 (Score) = 0.97 |
| | | | Micro F1 (Score) = 0.98 |
| X-ray Images | Tuagi et al. [170] | Chest X-ray Images [95] | Accuracy (%) = 96.45 |
| | Duong et al. [171] | Montgomery County (MC) CXR Images [183]<br>Shenzhen CXR Dataset [183]<br>COVID-19 Dataset [184] | Accuracy (%) = 97.92 |
| | Park et al. [174] | BIMCV COVID-19+ [96] | AUC = 0.949 |
| | | | Accuracy (%) = 86.8 |
| | | | Sensitivity (%) = 90.2 |
| | | | Specificity (%) = 86.2 |
| | Shome et al. [175] | COVID-19 Posterior-Anterior Chest Radiography Images [97]<br><br>Extensive COVID-19 X-Ray and CT Chest Images [98] | Accuracy (%) = 92 |
| | | | Precision (%) = 93 |
| | | | Recall (%) = 89 |
| | | | F1 Score = 0.91 |
| | | | AUC = 0.98 |
| | Rahha et al. [176] | Sars-CoV-2 [92] | Accuracy (%) = 94.62 |
| | | | Precision (%) = 96.77 |
| | | | Recall (%) = 96.77 |
| | | | Specificity (%) = 99.65 |
| | | | F1 Score = 0.967 |
| MRI Scans | Dai et al. [178] | MRNet Dataset [92] | AUC-ROC = 0.976 |
| | | | Accuracy (%) = 91.8 |
| | | | Sensitivity (%) = 96.8 |
| | | | Specificity (%) = 72.8 |
| OCT/Fundus Images | Kamran et al. [179] | Color Fundus Images [100] | Accuracy (%) = 85.7 |
| | | | Sensitivity (%) = 83.3 |
| | | | Specificity (%) = 90.0 |
| Histopathology Images | Zeid et al. [180] | Colorectal Cancer Histology Dataset [101] | Accuracy (%) = 93.3 |
| | | | Precision (%) = 93.33 |
| | | | Recall (%) = 93.44 |
| | | | F1 Score = 0.933 |

## 6.2. ViT in Region-based Lesion Detection

The proceeding section discusses the detection of anomalies in medical images using Vision Transformers, in relation to the aforementioned modalities.

### 6.2.1. Computed Tomography (CT) Scans:

*COVID-19 Detection* A framework for the detection of Covid-19 was proposed by Liang et al., with chest CT images as input [185]. The framework was composed of a CNN model for feature extraction, and then the SE attention module was integrated for the generation of attention vectors. Next, the transformer model was used to distinguish the features in the input. The study also proposed a method to resample the inputs, which also contributed to the efficiency of the model. The highest f1-score was 88.21 percent which was a 10 percent improvement from the baseline model. The dataset used, however, was small and imbalanced which doesnot validate the generalizability of the proposed model.

*Anomaly Detection* In the medical domain, various methodologies are proposed for the detection of anomalies. The authors of the paper [186], have proposed a transformer-based model, which was applied on various images i.e. retinal OCT, Head- CT Scans, and Brain- MRIs. The representation of the features was learned using autoencoders which were based on transformers. In addition, to detect the anomalies in multiscale, a transformer model was proposed with skip-connections, thus it reduced the usage of memory and cost of computation. The models were evaluated on AUROC, achieving 93 %, 95.81 %, and 98.38% for the datasets related to Head-CT, Brain- MRI, and retinal OCT, respectively. However, the proposed model still requires a further reduction in computational cost so that it can be used in real-time.

### 6.2.2. X-ray or Radiographic Images:

*COVID-19 Diagnosis* Since X-rays are comparatively cost-effective and a faster way of diagnosing the virus, several researchers have proposed methods for detection using chest x-rays.In the paper [187], another approach is proposed to detect covid-19 using chest X-rays. An adaptive attention network is used which consists of ResNet and an attention-based encoder. ResNet is used to learn the feature representations and the Attention module is then utilised for detection of the infectious areas. The proposed model was compared with different CNN models on three different datasets. The evaluation indicated that the proposed model performed significantly better. Moreover, the performance metrics included Accuracy, sensitivity, precision, and F1 score. The highest accuracy achieved by the model was 98.5

Similarly, To capture the global context, the authors Kumar et al. have used vision transformers on both X-ray images and CT images of the chest for the diagnosis of Covid-19 [188]. The data used was labelled as normal, pneumonia and covid-19. Furthermore, to address the issue of scarce data, transfer learning is used followed by explainability through visualisation of the infected areas. The proposed method was compared with other models i.e InceptionV3, CoroNet, CovidNet, etc. The results were evaluated using the metrics; precision, recall, f1-score, accuracy, and specificity. The proposed model outperformed the CNN models reaching the accuracy of 0.96 and 0.98 for CT scans and X-ray images, respectively. However, work on severity information requires attention in both [187] [188].

*Pulmonary Lesions Detection* In the initial assessment of lung cancer, one of the most used techniques is chest radiography. Since it is essential to diagnose cancer at an earlier stage, many methodologies have been proposed to detect pulmonary lesions. The study is presented by [189], it has proposed two architectures; convolution networks with attention feedback, and recurrent attention model with annotation feedback. The first method uses CNN to learn the features, and generate saliency maps as the soft attention mechanism was incorporated. Next, a recurrent attention model with attention feedback was proposed. The proposed architecture uses reinforcement learning for better performance of the model. The architectures were evaluated through precision, recall, f1-score, and accuracy. The highest accuracy achieved was 85 percent for classification, and 74 percent for localization. Thus, the architectures require improvement regarding the reduction of computation time and accuracy.

### 6.2.3. OCT or Fundus Images:

*Microaneurysms Detection* The early diagnosis of lesions in diabetic retinopathy can be done by the detection of microaneurysms (MA). Since it is difficult to locate them because of their size, several methodologies have been proposed. In this study [190], the proposed methodology for the detection of MAs comprises three stages. First, the images are preprocessed to improve the quality. Second, a deep network is used with an attention mechanism for detection. Third, the correspondence between Microaneaurysms and blood vessels is exploited for the final results.

The performance metrics used for evaluation were precision, recall, sensitivity, and f1-score. The proposed method outperformed prior proposed models with a sensitivity of 0.86. Nevertheless, the model was trained on the images from one type of camera, which does not validate the generalizability of the proposed methodology.

*Glaucoma Detection* The authors of the paper [191], have proposed a methodology for the detection of a disease known as Glaucoma. It causes the loss of vision and is irreversible. The paper has presented a CNN model which was attention-based for the detection of the disease. Furthermore, due to the attention module, the localized features were also visualized, giving results more explainability. The proposed architecture first locate the area and then classify the disease. The detection was evaluated using the performance metrics; accuracy, sensitivity, specificity, AUC, and, F2-score, with the highest accuracy achieved of 96.2 percent. However, in the network, the models may identify regions with useless information which may hinder the performance of the model.

Further, Xu et al. have presented a model which consists of an attention module along with transfer learning for the detection of glaucoma [192]. This work has contributed towards the discrimination of general and specific features. Since the models are not able to identify the regions that may give no information, the proposed methodology can extract the regions with more information. In addition, with the attention module, the regions can also be visualized. The model was then evaluated on two different datasets, achieving the highest accuracy, sensitivity, specificity, and AUC of 85.77 percent, 84.9 percent , 86.9 percent, and 0.929, respectively. Lastly, this method can be further validated, by applying it to various other eye diseases.

#### *6.2.4. Histopathology Images:*

*Cancer Detection* Barret's esophagus (BE) refers to the damaging of the swallowing tube that connects the mouth to the stomach because of acid reflux [193]. Ultimately, it increases the risk of esophagus cancer i.e. adenocarcinoma [194]. Moreover, patients that suffer from BE are at a higher risk of cancer. The detection of lesions at an early stage can prevent the suffering of patients from cancer, with a better survival rate.In the paper[195], attention-based deep neural networks were proposed for the detection of cancerous and precancerous esophagus tissues. The model uses attention-based mechanisms to detect the cancerous tissues belonging to the classes; normal, BE-no-dysplasia, BE-with-dysplasia, and adenocarcinoma. The mechanism does not require annotations for regions of interest, thus, it dynamically identifies the ROIs. Hence, it is independent of the annotated bounding box and does not require a fixed size of input images. The proposed method was compared with the sliding window approach based on the performance metrics; accuracy, recall, precision, and accuracy. The model outperformed the sliding window method in all classes with an average accuracy of 0.83. However, the model was trained on a small dataset, hence the robustness of the model still needs to be verified using more data.

The issues regarding whole-slide images in terms of detection, include poor adaptability of the model, explainability, and resource-efficient model. The authors of the paper[196], have proposed a model known as clustering-constrained-attention multiple-instance learning(CLAM). It was applied to detect three types of cancers; renal cell carcinoma, non-small cell lung cancer, and breast cancer lymph node metastasis. The proposed method CLAM is a weakly supervised algorithm, it uses an attention module to determine the regions, and classify the cancer type. In addition, it also localized the affected regions with interpretability. The models were evaluated using AUC, hence, it was greater than 0.95. On contrary to this data-efficient model, it considers various locations as independent, thus, leading to a less context-aware model.

Next, another model was used for the detection of cancer leading to the prediction of survival prediction [197]. The framework is a multimodal co-attention transformer( MCAT), that learns the correspondence between WSI's and genomic features. The attention module ensures interpretability along with the reduction of memory usage of image bags. The model was applied to five different cancer datasets, and the results were compared with the state-of-the-art models.

The table 5 given below summarized the performance gain by the reviewed articles of the detection category.

Table 5: List of datasets and performance measures employed by researchers for Region-based Lesion Detection.

| **Modality** | **Publication** | **Dataset** | **Performance Measures** |
|---|---|---|---|
| CT Scans | Mondal et al. [188] | COVIDx-CT-2A [104] | Accuracy (%) = 98.1 |
| | | | Recall(%)=96 |
| | | | Precision(%)=96 |
| | | | Specificity(%)=98.8 |
| | | | F1(Score)=0.96 |
| | Liang et al. [185] | COV19-CT-DB [93] | Macro F1 (Score) = 88.21 |
| | | | Micro F1 (Score) = 0.98 |
| X-rays | Lin et al. [187] | COVIDx [99] | Accuracy (%) = 95 |
| | | | sensitivity(%)=97 |
| | | | Precision(%)=98.98 |
| | | | Specificity(%)=99.47 |
| | | | F1(Score)=0.97 |
| | | COVIDGR-E [102] | Accuracy (%) = 89.53 |
| | | | sensitivity(%)=86.05 |
| | | | Precision(%)=83.15 |
| | | | Specificity(%)=91.28 |
| | | | F1(Score)=0.84 |
| | Lin et al. [187] | DLAI3 | Accuracy (%) = 98.55<br>sensitivity(%)=98.63 |
| | | | Precision(%)=98.63 |
| | | | Specificity(%)=99.90 |
| | | | F1(Score)=0.98 |
| | | | Precision (%) = 15 |
| | Pesce et al. [189] | A dataset consisting of 745,479 chest x-ray exams collected from the historical archives of Guy's and St. Thomas' NHS Foundation Trust in London from January 2005 to March 2016 | Sensitivity (%) = 65 |
| | | | Average Overlap (%) = 43 |
| Fundus Images | Zhang et al. [190] | IDRiD [103] | Accuracy (%) = 94.3 |
| | | | Precision (%) = 87.2 |
| | | | Recall (%) = 81.0 |
| | | | F1 Score = 0.840 |
| | | | Sensitivity (%) = 86.8 |
| | Li et al. [191] | LAG Database (obtained from Chinese Glaucoma Study Alliance (CGSA) and Beijing Tongren Hospital.) [191] | Accuracy (%) = 96.2 |
| | | | Sensitivity (%) = 95.4 |
| | | | Specificity (%) = 96.7 |
| | | | AUC = 0.983 |
| | | | F2 Score = 0.954 |
| | Xu et al. [192] | LAG Database (obtained from Chinese Glaucoma Study Alliance (CGSA) and Beijing Tongren Hospital.) [191] | Accuracy (%) = 85.7 |
| | | | Sensitivity (%) = 84.9 |
| | | | Specificity (%) = 86.9 |
| | | | AUC = 0.929 |
| | Tomita et al. [195] | | Accuracy (%) = 83.0 |
| | | | Recall (%) = 60.0 |
| Histopathology Images | | Histological images between January 1, 2016, and December 31, 2018, at Dartmouth-Hitchcock Medical Center (Lebanon, New Hampshire) were collected. | Precision (%) = 62.0 |
| | | | F1 Score = 0.59 |
| | Chen [197] | The Cancer Genome Atlas [105] | Concordance Index (c-Index) = 0.653 |

## 6.3. ViT in Anatomical Structure Segmentation

Clear cut and detailed segmentation is a decisive step in image guided treatment and computer-aided diagnosis. A great deal of image segmentation models have been proposed In the last 40 years from traditional models to deep neural networks. But since the emergence of transformers, they have outperformed all the state of art segmentation models. Transformers functions prominently in error free segmentation of medical images because of their capability to model the global context. As the organs lay out over a wide receptive field, hence, transformers can easily encode these organs by modeling the association of pixels that are distant spatially. Moreover, the background is dispersed in medical scans, for that reason gaining the understanding of the global context between those pixels that relate to the background will be beneficial for the model to do the unerring classification. Below we reviewed experiments that tried to exploit ViT based models for a faultless segmentation. We divided these experiments in accordance with different modalities used for medical imaging. In the end we give all the results obtained during these experiments on specific datasets in tabular form.

### *6.3.1. Computed Tomography (CT) Scans:*

*Coronary Artery Segmentation* A precise and correct segmentation of CAC is advantageous for early CVD diagnosis. But as CAC has blurry and distorted boundaries, the task of segmentation is not very much satisfactory. To address this issue Ning et al. [198] introduced an efficient multiscale vision Transformer for the segmentation of coronary artery calcium and named it as CAC- EMVT. This architecture utilized the local as well as global features and then used them collectively to model the long and short-term dependencies. Their model was comprised of three main modules, (a) KFS, Key Factor Sampling module that they utilized for extracting the key factors from the image. These key factors were made use for low-level reconstruction of highly structured features, (b) NSCF, a non local sparse net fusion module, that was used to model the information of high level features of texture, (c) NMCA, a non local multiscale context aggregation, that was used to get the dependencies of long range at different scales. Experimental results showed that their model outperformed the state of the art methods at that that by giving a Dice similarity co-efficient of 75.39% $\pm$ 3.17.

Lee et al. [199] introduced a new concept of template transformer networks for segmentation through shape priors (TETRIS), and performed coronary artery segmentation through their model. In this concept they used an end to end trainable Spatial Transformer (STN)[200] to deform a shape template to complement the under laying region of interest. They also used this concept of incorporating the priors to the state of the art CNN and U-Net used for binary classification. The experimental results of TETRIS and U-Net[139] incorporating the prior, were able to produce singly connected components because they were given the prior information and gave to dice scores of 0.787 and 0.854 respectively. They also compared the U-Net[139] with FCN by giving the prior shape but FCN[201] didn't perform that well giving the Dice Score of 0.79 only.

*Lung Tumor Segmentation* PET-CT segmentation requires information from both PET and CT modality. Most of the models get the segmentation information of these modalities separately. In a study, Fu et al. [202] established a module MSAM, Multi model Spatial Attention Module, a deep learning based framework for lung tumor segmentation in PET-CT. MSAM was an impulsive module and it was able to highlight the spatial areas or regions that are linked to the tumor and censored the normal regions of input spontaneously. This module was followed by a CNN that was acting as a backbone and this CNN was performing the segmentation task on the map that was provided to it as an input from the multi model spatial attention module. U-Net[139] was used as backbone for that purpose. They performed their experiments on two clinical PET-CT datasets of NSCLC and STS. The results of the experiments gave a dice similarity coefficient of 71.44% and 62.26% for NSCLC and STS datasets respectively. in order to refine this architecture, more better procedures can be used to further improve the segmentation.

*Renal Tumor Segmentation* Due to the diversification that is present in size and pose, the task of segmentation has become a strenuous task. Hence La et al. [203] proposed a network that was both size and pose invariant and they tested their network for renal tumor segmentation on 2D CT Scan images. Their architecture was comprised of three sequential modules that worked together in the training process. First was the regression module that they used to find the similarity matrix of input image to the ground truth. Second module was used to find the region of interest and they named it as differentiable module. The third and last module was used to perform the segmentation task and they used U-Net[139] for this purpose. They used the Spatial Transformer (STN)[200] in their architecture to automatically detect the bounding box which saved time. Results indicated that the training time was reduced by 8 hours and the Dice Score for kidney almost remained same which was 88.01%, but in case of renal tumor, the score got better from 85.52% to 87.12%. one of the shortcomings of their model was that it was valid for only small set of data.

*Aortic Valve Segmentation* Basic CNN models for segmentation were performing good on 2D images and they were struggling against 3D medical imaging. Hence Pak et al. [204] proposed a deep learning based architecture for the segmentation of 3D CT Scan images. This network was comprised of a baseline U-Net Architecture that performed the basic segmentation task and a Spatial Transformer[200] that was used to perform some affine transformation. The use of only U-Net[139] was not sufficient for the segmentation tasks as it requires a lot of memory and also result in decrease of accuracy. Hence they used a spatial Transformer (STN)[200] which reduced the size of input image by performing some transformation and hence it resulted in better computation. they utilize their model to perform aortic valve segmentation. Upon testing their model on different patients data, the Dice Score coefficient they get was 0.717.

*Bone Segmentation* In order to perform the segmentation of bone as well as the localization of the anatomical landmarks of cone beamed computed tomography data simultaneously Lian et al. [205] proposed a network called dynamic transformer network (DTNet). Their model contributed in three parts. In the first part, a synergic architecture was made to accurately catch the global context and fine grained details of image for volume to volume prediction. Secondly, by using the anatomic reliance between landmarks RDLs are made to collectively degenerate the large 3D heatmaps of every landmarks. Thirdly, ATMs are made for the compliant learning of context specific feature embedding from mutual feature bases. While doing the experiments on CT scans of mandible, the segmentation DSC came out to be 93.95% with a std dev of 1.30 whereas for localization the RMSE was $1.95 \pm 0.43$. these results were better than U-Net and other models that were used for comparison.

*Lesion Segmentation* The early diagnosis of AIS provides valuable knowledge about the disease. But for a human eye it is burdensome to discriminate delicate changes in pathology. Hence, Luo et al. Luo et al. [206] proposed a network for the segmentation of Acute Ischemic Stroke (AIS) that was based on self-attention. This mechanism had an encoder and a decoder. The encoder was comprised of a CNN as a backbone and a transformer. This encoder part picked the global context features. The decoder part consisted of Multi Head Cross Attention (MHCA) module which up sampled the feature maps that were coming from the encoder. These feature maps were connected via skip connections. The backbone CNN used was RESNET-50. Their experimental results were compared to attention U-Net[207], U-Net[139] and TransUNet[208] but their model outperformed them by giving the Dice Similarity Coefficient of segmentation of lesion up to 73.58% which was better than all other compared models.

*Segmentation of Organs* Although transformers help in capturing the long term dependencies but when it comes to the segmentation of 3D images, the dependencies face extreme computation. Hence to reduce some computations, Xie et al. [209] presented CoTr, which is a combination of convolutional network and transformer. Instead of using simple transformer, they introduced deformable transfers for catching the long range dependencies (DeTrans). DeTrans focusses on only a few key points, which greatly reduces the computational complexity, which also allows to process multi scale images, which are quite important to attain an accurate segmentation. they tested their model on BCV dataset that includes the images of 11 different human organs. CoTr achieved a Dice Score of 85 and Hausdorff Distance of 4.01 on avg and these measures were better than other methods that they used to compare their model. Further enhancement in their model could be that it can be enhanced by extended it to operate on different modalities.

#### 6.3.2. Magnetic Resonance Imaging (MRI) Scans:

*Brain Tumor Segmentation* Glioma segmentation and prediction of IDH genotyping is an important and difficult task due to similarities present in intra and inter-tumor. To address this problem Cheng et al. Cheng et al.[210] proposed an MRI based fully automated multi model that could predict IDH genotyping and Glioma segmentation at the same time. The pre existing methods were not able to perform the both tasks at the same time, also these methods faced the problems of inter and intra-tumour heterogeneity. So they address these issues by using a joint CNN and Transformer encoder. The transformer was used to extract the global features that were used for the glioma segmentation. It also contained a multi-scale classifier, which was used for IDH genotyping. A multi-task loss was then used to balance the segmentation and IDH genotyping and this loss collectively joined the classification loss and segmentation loss. In the end they proposed an unpredictability aware pseudo-label-selection to make pseudo-labels for IDH on a large unlabeled Dataset. They named their model as MTTU-Net. On experiments their model improved the HD95 and Dice score 1.69mm and 1.23% for glioma segmentation and 2.13% and 4.28% in case of AUC and accuracy respectively.

Sagar et al. [211] proposed ViTBIS, vision transformer for bio medical image segmentation, that was based on encoder and decoder architecture. Both encoder and decoder had transformer inside them. The feauture map of input image was split into three different convolutions before it was fed to the transformer. These convolutions were, 1X1, 3X3 and 5X5. These three different feature maps were concatenated with the help of concat operator, then it was fed to the transformer in the encoder. These transformer had the attention mechanism inside them the transformers of encoder and decoder were joined together via skip connections. The same architecture of multiscale was used in the decoder as well. Before producing a segmentation mask after linear projection, different sizes were concatenated via concat operator. Upon testing their architecture on a public dataset for brain tumor segmentation the DSC achieved was 0.86 which was better than other state of the art CNN and transformer networks.

*Brain Tissue Segmentation* In order to solve the problem of multi model medical image segmentation, Sun et al. [212] presented a novel multi model architecture based on transformer and Convolutional Neural Network for multi model image segmentation and named it as HybridCTrm, and used this model to segment different brain tissues. This network used two paths for the image encoding, one path was from the CNN and the other path was from Transformer. Then the representation of image from both paths were joined together for decoding and the segmentation purpose. The CNN controlled the rapid convergence of gradient descent while extracting the local features, whereas the non local features were extracted by the transformer. They used two strategies for the fusion, one was the single path strategy and the other was the multiple path strategy and used both of these strategies in their experiments. Experiments were carried out on two different datasets and by following both strategies. On MRBrainS dataset the DSC came out to be 82.98 and 83.47 for sigle and multiple path strategies respectively whereas on the iSeg-2017 dataset these scores were 86.75 and 87.16 which were better than the models they used to do the comparison like HyperDenseNet[213],

*Brain Structure Segmentation* A good deal of deep learning architectures used to perform the task of segmentation on medical images confront the problem of noise at the inference time and result in inaccurate result. To address this problem Sinclair et al. [214] proposed a network, Atlas-ISTN, atlas image and spatial transformer network, that was able to perform both registration as well as segmentation on 2D and 3D data of Brain structure. This network could perform segmentation on numerous interest regions/ structures and to register the atlas label map to an in-between segmentation(pixel-wise). This model was also able to do the fine tuning of the parameters at the inference time in order to achieve better pixel wise segmentation, due to which the effect of noise in the image also reduced. This model was then tested on three different datasets, two 3D and one 2D. the results were compared with U-Net and this model was performing better than U-Net[139] giving a DSC of 0.888.

*Cardiac Segmentation* Combining the shared information of any organ from different modalities is very helpful for learning and multi modality processing. In order to schieve this, Chartsias et al. [215] proposed disentangled, align and fuse network, DAFNet, that was able to learn the information present in different modalities input, hence producing a more precise segmentation mask. Anatomical factors from different inputs are combined and processed at the same time. DAFNet collected the information present in different modalities despite of the fact that few labels(supervised) are there or even no labels (unsupervised). Spatial Transformer was used to align the anatomical factors in case of image misregistration. They evaluated their model by performing L2, T1 and Cardiac segmentation on different datasets. Their model was able to perform on both single model and multi model inputs and it outperformed other models when it was trained on single modality input whether with few labels(semi supervised) or no labels (unsupervised).

Defining the right ventricle (RV) structure in cardiac segmentation is a stretching work to do because of its complex and multiplex structure. Hence it requires short axis as well as long axis images. In order to address this issue Gao et al. [216] established a consistency based co training mechanism that used the geometric relationships between different view CMR images for the segmentation. Along with this mechanism, they also used the U-Net[139] architecture in order to capture some long range dependencies. Evaluation of the model was done on the M&MS-2 challenge data set and the Dice score came out to be 0.83 and 0.86 for short axis and long axis respectively.

*Colerectal Cancer Segmentation* In a study done to segment the colorectal cancer region Sui et al. [217] established a novel approach, based on transformer, that performed the segmentation as well as detection of colorectal cancer region collectively. Their model was based on two pipelines, one for the detection and the other for the segmentation. In the detection part, region proposals were generated. They utilize image level decision approach that was based on auto encoders. Whereas in the segmentation part they used patches of the image as input and to make the final mask prediction, class embeddings were used. They compared their model with the Faster CNN and Yolo-v3 for the detection task and their model performed exceptionally well on the used dataset, giving an accuracy of 88.6% where as the segmentation score came out to be 91.1% which was way better than U-Net[139] and FCN[201].

### 6.3.3. X-ray or Radiographic Images:

*Breast Tumor Segmentation* Correct and accurate segmentation of tumor in ABVS is a difficult task because the size of image is huge and its quality is low. In order to segment tumor from the ABVS images, Liu et al. [218] adopted the use of both transformers and CNNs and named their model as 3D-UNet. They joined the attention module and the U-Net[139] model. For further improvements in the performance they also made use of Atrous Spatial Pyramid Pooling (ASPP) in their model. ASPP can help catch the information at multi scales. They compared their model with different 3D segmentation models like 3D FCN[219], 3D PSPNet[220] and recorded the Dice Score Coefficient. Their score recorded as 76.36±6.11, which was better than the networks that were used for the comparison.

*Anatomical Structure Segmentation* Most of the segmentation networks work on supervised learning where expert labeled image is required as a label and this is an obstacle if there aren't much experts available. Hence in order to make an annotation efficient Lu et al. [221] introduced Contour Transformer Network, CTN, which is an annotation efficient segmentation method for anatomical structures. They copied the human ability doing the segmentation of anatomical structures with very less exemplars available. To achieve this, they proposed a semi supervised learning mechanism that utilize the resemblance of structure and appearance of the desired object between unlabeled and labeled images. They made the segmentations of anatomy in the form of contour evolution process and model the behavior by GCNs. They named their model as one shot anatomy segmentation model. On performing the segmentation on four different anatomies, their model comprehensively performed better than u supervised learning mechanisms and performed competitively against the supervised state of the art methods. Upon experiments the accuracy of one shot model came out to be 96.58% which was almost 15% better than Braintorm, which is another one shot based model. Whereas in comparison with non-learning based model, the accuracy was 16% improved. one shortcoming of their network is that their network only performed on 2D data. hence extending this architecture to work on 3D data would be an important step in the field of 3D segmentation.

*Guide Wire Segmentation* A study done by researchers tried to resolve the task of segmentation of guide wire in X-ray fluoroscopy sequence. Zhang et al. [222] proposed a network that takes in account current frame as well as the previous frame while taking input for the guide-wire segmentation. By considering both frames helped them in obtaining the temporary information. Their network contained two parts, one was a CNN and the other was a transformer. The CNN wasn't able to capture the global features hence transformer came into play that can learn the global features by using its attention mechanism. CNN and transformer lied in the encoder part whereas decoder contained up sampling, concatenating operations and convolutions. They evaluated their model on datasets from three different hospitals and measured the F1-score and compared their score with other state of the art models like Frrnet[223], Parnet[224] and U-Net[139] and their model was outperforming all other models.

*Tooth Root Segmentation* An accurate and precise segmentation of the boundaries present in the roots of the tooth is necessary to attain a perfect root canal therapy assessment. Li et al. [225] introduced a Group Transformer Network (GT U-Net) in order to achieve segmentation of root boundaries. Their model's structure was similar to the U-Net but they used group of transformers in place of encoders and decoders. Also in order to incorporate the prior knowledge they used Fourier Descriptor Loss. Their model achieved an accuracy of 96.31% and f1 score of 84.58% outperforming other state of the art models.

#### 6.3.4. OCT or Fundus Images:

*Drusen Segmentation* It is very crucial to diagnose the AMD at an early stage in retinal OCT images via Drusen Segmentation. In order to achieve an accurate segmentation Wang et al. [226] proposed a multi scale transformer global attention network MsTGANet for the segmentation of drusen in retinal OCT images. Their model was composed of a U-shaped architecture containing an encoder and decoder. To collect the non-local features at different scales with ling term dependencies from multiple encoder layers, a novel multi scale transformer non local module is proposed and used at encoder's top. Another module, MsGCS was introduced to assist the model to join different semantic knowledge between encoder and decoder. They also introduced a semi supervised version of MsTGANet. This version was comprised of pseudo-labeled data augmentation strategy. This model can used huge amount of un labeled data in order to increase the performance on segmentation, upon experiments the DSC came out to be 0.8692 with a std of 0.0052 outperforming the other state of the art models. this model was trained on a smaller dataset, however it will be better to collect a larger set of data in order to see its efficiency. Also, different semi-supervised learning approaches can also be used to further improve its performance.

The table 6 given below summarized the performance gain by the reviewed articles of the segmentation category.

Table 6: A list of datasets and performance measures adopted by researchers for Segmentation.

| **Modality** | **Publication** | **Dataset** | **Performance Measures** |
|---|---|---|---|
| MRI | Cheng et al.[210] | BRaTS2020[107] | Dice Score = 0.90 |
| | | | Hausdorff Distance = 4.4 |
| | | | AUC(%) = 91.04 |
| | | | Accuracy(%) = 90 |
| | | | Sensitivity(%) = 87.50 |
| | | | Specificity(%) = 92.11 |
| | Chartasis et al.[215] | ERI[110] | Dice Score = 0.82 |
| | | CHAOS[111] | Dice Score = 0.85 |
| | Sun et al.[212] | MRBrainS[108] | Dice Score = 0.83 |
| | | iSeg2017[106] | Dice Score = 0.87 |
| | Sinclair et al.[214] | UKBB[109] | Dice Score = 0.86 |
| | | | Hausdorff Distance = 7.2 |
| | Sagar et al.[211] | BRaTS2019 | Dice Score = 0.86 |
| | | | Hausdorff Distance = 7.1 |
| | Gao et al.[216] | M&MS2 | Dice Score = 0.86 |
| | | | Hausdorff Distance = 9.6 |
| X-ray Images | Lu et al.[221] | OAI | IOU(%) = 97.32 |
| | | | Hausdorff Distance = 6.0 |
| | | JSRT[113] | IOU(%) = 94.75 |
| | | | Hausdorff Distance = 12.1 |
| | Li et al. [225] | DRIVE[112] | Dice Score = 0.92 |
| CT Scans | La et al.[203] | KiTS2019[114] | Dice Score = 0.88 |
| OCT/Fundus Images | Wang et al.[226] | USCD[115] | Dice Score = 0.86 |

## 6.4. Clinical Report Generation

In this section, we briefly describe various transformer models to generate the medical reports and address the preceding challenges associated with automatic clinical report generation.

### 6.4.1. *Supervised Learning Based Approaches*

Supervised learning refers to a type of learning algorithms that learn under the presence of a supervisor. An input from the training set is passed through the network then the output of the network is compared to the desired output and learning weights are updated accordingly. Following studies have employed supervised learning in their methodologies.

*Incorporating Global Level Features* Global level features are extracted from the entire medical image i.e. encoded features of both normal and disease regions in the image. Following studies have incorporated this notion into their methodologies. You et al. [144] proposed a transformer-based architecture, AlignTransformer. They resolved the data bias and long sequence modeling problems to generate a coherent medical report by delineating the normal and abnormal regions. They used ResNet-50 pre-trained on ImageNet and fine-tuned on CheXpert dataset to extract visual features. Furthermore, they fed the extracted visual features into the pre-trained multi-label classification netwrok to predict the disease tags. Align hierarchical attention as an encoder aligned the disease tags and visual regions by learning the correlation and relationship between them. Moreover, they acquired the multi-grained disease-grounded visual features from the aligned disease tags and visual regions to alleviate the data bias problem. Multi-grained transformer as a decoder exploited the multi-grained disease grounded visual features to generate a proper medical report. In automatic evaluation, they compared their experimental results with the previous state-of-the-art models i.e. R2Gen, PPKED,and SentSAT +KG etc. and achieved competitive results on IU X-ray and MIMIC-CXR datasets. In human evaluation, the results of their model were far better than that of R2Gen model. Similarly, Amjoud et al. [227] also proposed a transformer-based deep learning model for generating long and detailed reports of chest x-ray images. They used a pre-trained DenseNet-121 [82] instead of ResNet-50 [144] to avoid gradient vanishing problem and redundant feature maps. They suppressed the last classification layer of the pre-trained model to extract global and regional features from medical images. After that, the extracted features were fed as input into the encoder to map them into a sequence of continuous representations. They modified the decoder of the vanilla transformer by adding a relational memory module to the normalization layer. Experiments demonstrated that their model generated detailed findings reports for IU chest x-rays test images and outperformed the state-of-the-art models for BLUE-1, BLUE-2, and ROUGE metrics with 0.479, 0.359, and 0.380 scores respectively. However, the model could not perform well for BLUE-3, BLUE-4, and METEOR metrics. Also, they used a small corpus for training, as a result, some sentences were unseen during inference which lead to the scattering problem.

In another work, Pahwa et al. [143] leveraged the skip connections by proposing a transformer-based architecture named MEDSKIP by modifying a high-resolution network (HRNet). They modified (HRNet) [228] for visual features extraction by incorporating skip connections along with convolutional block attention modules (CBAM). First, they extracted features representation from each down-sampled layer and after extracting crucial features using attention, CBAM concatenated them. CBAM constituted spatially and channel attention sub-modules for inferring a 1D channel attention map and a 2D spatial attention map respectively. The proposed architecture also contained a memory-driven transformer which constituted a standard encoder but the decoder contained a memory-driven conditional normalization layer to incorporate relational memory. The decoder facilitated the learning from patterns in reports and recorded key information of the generated process. Extensive experiments on two publicly available datasets PEIR GROSS and IU chest x-rays showed that their proposed model had given the state-of-the-art results for BLEU, METEOR, and ROGUE metrics.

*Incorporating Global and Local Level Features* A medical image contains both normal and disease regions. To encode the disease regions of the image, previous studies encoded the complete image which lead to the encoding of irrelevant visual content which is adverse for radiology report generation. Some diseases have strong correlation and finding those correlation is beneficial for generating report for rare diseases. Various studies tried to take advantage of this feature by constructing correlation matrix in the encoding stage by data-driven methodologies or expert knowledge but these studies failed to decode these correlations effectively while decoding.

To address these problems, Jia et al. [229] leveraged the transformer-based architecture and proposed a few-shot radiology report generation model, namely TransGen. In the encoding stage, they introduced a semantic-aware visual Learning (SVL) module in which they used ResNet101 to identify and capture the disease regions of rare diseases. They captured the disease regions from the image itself and the feature map generated at time step (t-1) by learning the two masks respectively to refine the visual representation of rare diseases. They adopted a weighted sum of these two masks at time step t to learn the visual representations efficiently by incorporating both global and local level information. For efficient decoding of encoded correlation among the diseases, the memory augmented semantic enhancement module was introduced at the decoding stage. Experiments demonstrated that their model outperformed the state-of-art models on the MIMIC-CXR dataset but could not perform well for the IU X-ray dataset.

Similarly, Lee et al.[230] also incorporated both local and global level features by proposing Cross Encoder-Decoder Transformer (CEDT) contained a Global-Local Visual Extractor. They used a convolution neural network (e.g.ResNet101) as a global visual extractor to encode the complete radiology image into a sequence of patch features to accurately capture the features at the global level i.e. bone structure or size of the organ. However, while incorporating global-level features, it was difficult to encode the exact location and the size of the lesion area. To address this problem, they cropped the disease regions of the image with the last layer of the CNN using the attention-guided mask inference process and after resizing to the same size as the image, used them as input to the local feature extractor to extract the local level features. Then, they concatenated the local visual features and global visual features and used them as input to the CEDT. The standard transformer uses only the last layer information but they [230] also used low-level features in addition to the high-level features by using the concept of [231]. They used multiple encoders to get the all-level information from them and utilize the outputs of all encoders on each decoder using parallel multi-headed attention. They added the extracted features of each encoder layer which resulted in better captioning than the baseline model R2Gen. Furthermore, they also employed MCLN and RM for recording and utilizing the important information. Extensive experiments demonstrated that their model outperformed the baseline model for BLEU-1, BLEU-2, BLEU-3, METEOR, and ROUGE-L on IU X-ray dataset. They also performed experiment with pre-trained GLVE but it could not perform well for the BLEU-4 metric.

### 6.4.2. Reinforcement Learning Based Approach

Previous studies [143, 144, 227, 229, 230] have used supervised learning approaches to generate medical reports. Supervised learning approaches are prone to exposure bias problems in language modeling methods. To address this problem, Xiong et al. [232] proposed a novel hierarchical neural network architecture using reinforcement learning to generate a long coherent medical report. They incorporated the self-critical reinforcement learning method into the detector, encoder, and captioning decoder. Previous studies used only top-down visual encoders, however, this was the first study that incorporated a bottom-up visual detector as well to extract semantic rich features from the medical images. For this purpose, firstly they used DenseNet-121, pre-trained on chest X-ray 14 dataset, to detect the region of interest (ROI) proposals using a bottom-up attention mechanism. The region detector outputted a set of ROI proposals along with classified classes and some associated attributes. Secondly, they used top-down transformer visual encoder to extract further pixel-wise visual information from proposed ROI using pooling operations. Lastly, the transformer captioning module used improved ROI proposals as input from the transformer visual encoder and generated descriptive sentences for each proposed ROI by calculating reward directly using the CIDEr metric. Their proposed architecture outperformed the state-of-the-art methods for the CIDEr evaluation metric on the IU X-ray dataset but for the BELU-1 metric, their model could not perform state-of-the-art. Their model over-fitted as they used only the findings portion of the generated medical report. This problem can be resolved using a larger labelled dataset.

The table 7 given below summarized the performance gain by the reviewed articles of the clinical report generation.

Table 7: A list of datasets and performance measures adopted by researchers for Clinical Report Generation.

| **Modality** | **Publication** | **Dataset** | **Performance Measures** |
|---|---|---|---|
| X-ray Images | You et al. [144] | IU X-ray [116] | BLEU-1 = 0.484 |
| | | | BLEU-2 = 0.313 |
| | | | BLEU-3 = 0.225 |
| | | | BLEU-4=0.173 |
| | | | METERO=0.204 |
| | | | ROUGE-L=0.379 |
| | | MIMIC-CXR [117] | BLEU-1=0.378 |
| | | | BLEU-2=0.235 |
| | | | BLEU-3=0.156 |
| | | | BLEU-4=0.112<br>METERO=0.158 |
| | | | ROUGE-L = 0.283 |
| | Amjoud et al.[227] | IU X-ray [116] | BLEU-1=0.479 |
| | | | BLEU-2=0.359 |
| | | | ROUGE-L = 0.380 |
| X-rays | Pahwa et al. [143] | IU X-ray [116] | BLEU-1= 0.467 |
| | | | BLEU-2=0.297 |
| | | | BLEU-3=0.214 |
| | | | BLEU-4=0.162 |
| | | | METERO=0.187 |
| | | | ROUGE-L=0.355 |
| | | PEIR GROSS [118] | BLEU-1= 0.399 |
| | | | BLEU-2=0.278 |
| | | | BLEU-3=0.209 |
| | | | BLEU-4=0.148 |
| | | | METERO=0.176 |
| | | | ROUGE-L=0.414 |
| | Jia et al.[229] | IU X-ray [116] | BLEU-1=0.461 |
| | | | BLEU-2=0.285 |
| | | | BLEU-3=0.196 |
| | | | BLEU-4=0.145 |
| | | | ROUGE-L= 0.367 |
| | | | KA (%) = 0.367 |
| | | MIMIC-CXR [117] | BLEU-1=0.368 |
| | | | BLEU-2=0.243 |
| | | | BLEU-3=0.178 |
| X-ray Images | | | BLEU-4=0.138 |
| | | | ROUGE-L= 0.338 |
| | Lee et al.[230] | IU X-ray [116] | BLEU-1=0.5064 |
| | | | BLEU-2=0.3195 |
| | | | BLEU-3=0.2201 |
| | | | BLEU-4=0.1924 |
| | | | ROUGE-L= 0.3802 |
| | Xiong et al. [232] | X-ray [116] | BLEU-1=0.350 |
| | | | BLEU-2=0.234 |
| | | | BLEU-3=0.143 |
| | | | BLEU-4=0.096 |
| | | | CIDEr = 0.323 |

### 6.5. Miscellaneous ViT Applications in Medical Imaging

Tranformer-based architecture has also played a vital role in other applications of medical field i.e. in image synthesis, denoising the low dose computed tomography, and positron emission tomography images, enhancing the resolution of medical images etc.

#### 6.5.1. *functional Magnetic Resonance Imaging (fMRI) Scans:*

*Visualizing Regenerated Neural Visual Content* In the past decades, few studies have been conducted to decode the human brain neural activities into natural language sentences. The main purpose of decoding brain neural activity is basically to know the human brain's perception of textual or visual content. In the past, most deep learning studies focused on different task specific decodings, i.e. detection, classification, recognition etc., using functional magnetic resonance imaging (fMRI) data. With the advancement in technology, several research has been done in language decoding to decode the human brain semantics evoked by linguistic stimuli into natural language words or sentences. Inspiring from these tasks, Zhang et al. [59] proposed a hybrid language decoding model: CNN-Transformer to decode the visual stimuli evoked at multi-times by natural images into descriptive sentences. They exploited the concept of neural machine translation (NMT) [17] but the difference was in source sequence i.e. natural images in NMT but visual neural activities in [59]. To achieve this task, firstly they extracted meaningful semantic low-dimensional features from high-dimensional visual neural activities (low-level raw fMRI data) using two-layer one dimensional CNN. Secondly, the encoder part of the transformer encoded the semantic features into multi-level abstract representation. Lastly, the decoder of the transformer decoded the multi-level representation into descriptive natural language sentences. They compared their model with other decoding models and achieved state-of-the-art results for BLEU, CIDEr, and ROUGE metrics with 0.17, 0.66, and 0.18 scores respectively. In future, this transformer-based brain decoding technology will be useful for the people who are unable to transmit their visual perception into speech and will also be a breakthrough for neuro-scientists in understanding and decoding the neural activities of human brain.

#### *6.5.2. PET-CT Scans:*

*Medical Image Enhancement* Computed tomography is a non-invasive imaging technique for medical diagnosis. Since high exposure to X-rays radiation is deadly for humans and has become the main concern for medical practitioners. To lessen this effect of X-rays radiation, it is used in less quantity in CT scans but it poses some serious problems, i.e. less contrast, sharp features, corners, edges, and stronger noise, which affects the quality of CT scans. Although low-dose computed tomography (LDCT) is mainstream in clinical applications, but the posed problems cause hindrance in an effective clinical diagnosis. Many traditional methods (iterative methods) and convolution-based deep learning approaches were employed to acquire high quality LDCT images by deblurring and suppressing the artifacts. The high-frequency sub-band of images are noisy areas while the low-frequency sub-band are noise-free areas containing main image content. Since convolution-based methods are limited to extracting features from the local areas of images due to limited receptive fields. Therefore, transformers came into the scientific field and revolutionized the world with their facts of capturing long-range dependencies between image regions.

Keeping in account all these observations, Zhang et al. [233] proposed a transformer-based architecture to denoise the LDCT images by decomposing them into high frequency (HF) and low-frequency parts. Hence, the noise was only retained in the HF part and it also contained a plethora of image textures. To ensure the relationship between HF and LF parts of the image, they denoised the noisy HF part with the assistance of the latent texture of the LF part. For this purpose, they employed CNNs to extract corresponding texture and content features from the LF image part. Furthermore, they acquired the high-level features from the transformer using the texture features from the noisy LF and embeddings from the HF part. They used a modified transformer with three encoders and three decoders. Finally, they reconstructed the high-quality LDCT image piecewise by combining these high-level features with the content features from the LF part of the image. ent features from the LF part of the image. Extensive experiments demonstrated that their model outperformed all the baseline methods achieving 93.7% for VIT metrics, improved structure similarity by 12.3%, and root mean square error lowered by 40.5% on Mayo low-dose computed tomography images dataset. Since convolution-based methods cannot capture global contextual information, Wang et al. [6] first time proposed a convolution-free token-to-token vision transformer-based dilation network to denoise the LDCT images. They captured the noise from input medical images by learning deep features, after that, they remove the noisy estimated residual images in order to clean them. Firstly, they used tokenization block to tokenize the feature map patches into tokens. Secondly, they fed those tokens into transformer block, further, for enhancing the tokenization, they applied tokenization in cascaded form in token-to-token block. They further enlarged the receptive field and refined the contextual information using dilation in tokenization procedure. They performed dilation using reshaping, soft split and cyclic shift to enhance the context. They compared their model with other state-of-the-art models and their model outperformed for SSIM and RMSE metrics with 0.9144 and 8.7681 scores respectively, in denoising the images. Without down-scaling tokenization of the image can be enhanced.

PET/MRI can concurrently provide anatomical and morphological imaging information that aids in clinical disease diagnosis. PET acquires metabolic imaging information with the help of radio-tracers while MRI uses magnetic field gradients and radio waves to acquire images of soft body tissues. Although, these imaging modalities have applications in disease diagnosis, i.e. cancer, tumor, and brain diseases, but also pose some serious concerns. Since time requirement for PET imaging acquisition is high and as a result, patient discomfort can affect the image quality i.e. low contrast-to-noise ratio. The Information from MRI can assist in denoising the PET images using registration approach. Many traditional deep learning and computer vision methods proposed to enhance the PET image quality using MRI but due to discrepancy in modalities they could not extract contextual and spatial information efficiently. To address these problems. Zhang et al. [234] proposed a spatial adaptive and transformer fusion network (STFNet) for denoising low count PET with MRI. They adapted dual path using the spatial-adaptive block to extract features. For the fusion of high-level features, they modified the traditional transformer encoder and incorporated global attention to form a pixel-to-pixel relationship between MRI and PET. The fused feature map was used as input to the decoder for PET denoising. Their model obtained promising results for on RMSE,PSNR,SSIM, and PCC metrics.

Similarly, Luo et al. [235] proposed 3D Transformer-GAN to build a standard dose PET image from the low-dose PET image. They leveraged the CNN-Transformer architecture to incorporate both global and local information. CNN-based-Encoder extracted enriched spatial information from the input medical image. Moreover, the transformer captured the long-range dependency from the extracted features from the CNN-based-Encoder. The learned representation from the transformer are incorporated into the CNN-based-Decoder to restore them and reconstructed the standard PET image. Extensive experiments demonstrated that their model outperformed the state-of-the-art on real human brain dataset.

### 6.5.3. Magnetic Resonance Imaging (MRI) Scans:

*Medical Image Reconstruction* MRI is a prevalent non-invasive imaging technique but its acquisition process is slow. Consequently, there is a need to develop accelerated MRI methods. Simultaneously, several studies on deep neural networks have been conducted to develop state-of-the-art methods for accelerated MRI. Therefore, Korkmaz et al. [3] accelerated MRI by reconstructing full-sampled MRI images using unsupervised learning incorporating deep image prior framework to alleviate the problem of under-sampled acquisition. They proposed generative vision transformer based unsupervised MRI reconstruction architecture to increase the receptive field. Firstly, they performed generative non-linear mapping over latent and noisy space to improve the invertibility of the model. Secondly, they used cross-attention to improve the context, i.e. both global and local context, of image and latent features. They did not use any pre-trained model. Lastly, they performed inference on each individual subject to increase the generalization. They performed extensive experiments on accelerated multi-contrast brain MRI dataset and outperformed the convolutional-based generative models for PSNR and SSIM metrics.

Wang et al. [236] proposed a super-resolution approach to reconstruct the high resolution MRI scans from low resolution scans. They proposed adjacent slices feature transformer (ASFT) network. Firstly, they incorporated extra slices in the consecutive in-plane slices of an-isotropic 3D MRI images. Secondly, to harness the similarity between the consecutive slices they introduced a multi-branch features transformation and extraction (MFTE) block. Thirdly, to enrich the target high resolution slices with the information from the low resolution reference slices they filtered out the useless information using MFTE block. Moreover, they used spatial adaptive block to refine the features spatially. They used channel attention to incorporate the multi-scale features and consequently, they enhanced the super resolution. Their model achieved the state-of-the-art performance for super-resolution task.

*Medical Image Synthesis* Tissue morphology information acquired from multimodal medical images play an important role in the clinical practice. However, it is not commonly used because of the expensive capturing of these information. Generative models such as generative adversarial network (GAN) are in practice to artificially synthesis these images. GAN is a CNN based architecture that shows locality bias and spatial invariance across all the positions. Hu et al. [237] introduced a transformer based double-scale deep learning architecture for cross-modal medical image synthesis to incorporate log-range dependencies Double-scale GAN showed efficient performance on benchmark IXI MRI dataset.

*Medical Image Registration* Tissue deformation in a highly deformable anatomy is estimated using image registration. Diffeomorphic registration is one of the image registration techniques that preserves the invertible and one to one mapping between images. Current deep learning techniques lack the ability to handle long range relevance thus limiting the meaningful contextual correspondence between images. In the paper [238] a dual transformer network (DTN) model is proposed for the diffeomorphic registration of MR images. DTN, using self-attention mechanisms, model the relevance from both the separate and concatenated images embeddings, which facilitate contextual correspondence between anatomies. DTN consists of learnable embedding module, relevance learning module and registration field inference module. Diffeomorphic registration field is estimated using moving fixed and moving images for one-to-one mapping. DTN has two branches to learn the relevance based on the embeddings of separate one-channel images and concatenated two-channel images. First the Low-level image feature for the concatenated and separate images are extracted using CNN. Second, the image features, converted into sequences, are fed to DTN for feature enhancement, based on global correspondence. Concatenated features form both branches are then used to infer the velocity field and registration filed. The deformation field is represented as the exponential of the velocity which ensure the invertible mapping. Metric space is used to optimize the proposed DTN in an unsupervised manner.

#### 6.5.4. Endoscopy Images

*Medical Image Reconstruction* Endoscope is a minimally invasive medical imaging modality. It assists in medical diagnoses by acquiring accurate images of the internal organs of a patient's body. However, the small imaging sensor in the endoscope causes problems in acquiring the high magnified blood vessels images. Many traditional methods incorporating interpolated methods and deep learning have applications in reconstruction of super-resolution of single images. Mainly convolution-based deep learning methods are there in reconstructing high-resolution images which have problems in capturing the global context. Consequently, Gu et al. [239] leveraged transformers and proposed hybrid architecture with the convolutional neural network to enhance the texture of blood vessels. Firstly, CNN extracted the low-level features from low-resolution (LR) images and the transformer sampled the LR image into three different samples to extract texture from the image. Secondly, similarity was examined between different features extracted from transformer-based extractor and CNN-based extractor. Thirdly, they employed a texture migration method to interchange the information between multi-scale features extracted from the transformer and CNN to synthesize the image. Lastly, sub-pixel convolution operation was performed on migrated basic images to synthesize a high-resolution. Their model acquired promising qualitative and quantitative results than the CNN based single image super-resolution methods.

#### 6.5.5. Fluorescence Microscopy

*Quantitative Characterization of Anatomical Structure using Combination of Markers in Bone Marrow Vasculature and Fetal Liver Tissues* Fluorescence microscopy is a variant of traditional microscopy, which uses a higher intensity light source that excites a fluorescent species in a sample of interest. It is used for different purposes such as detailed examination of anatomical landmarks, cells, and cellular networks. Alvaro Gomariz et al. [240] proposed a Marker Sampling and Excite network to exploit the potential of attention-based network on the fluorescence microscopy datasets which are underexplored by the deep learning. The capability of the network is tested by the quantitative characterization in various datasets of microvessels in fetal liver tissues and bone marrow vasculature in 3D confocal microscopy datasets. Proposed model gives a convincing performance with F1 Score of 91.2% for sinusoids and 71.2% for arteries in the liver vasculature dataset.

*Denoising of Celullar Microscopy Images for Assisted Augmented Microscopy* Deep learning has greatly assisted augmented microscopy that enable high-quality microscopy images with using costly microscopy apparatus. Zhengyang Wang et al. [241] proposed global voxel transformer networks (GVTNets) that uses attention operators to address the limitations in already existing U-Net based neural. Instead of local operator that lack dynamic non-local information aggregation they used attention operators that allow global receptive field during prediction. They measured the performance of the model on three existing datasets for three different augmented microscopy tasks.

#### 6.5.6. Histopathology Images

Whole Slide Images contain rich content about the anatomical and morphological characteristics. To better depict the image content, pathologists frequently examine the tags associated with these images. Since this tagging process is labour-intensive, consequently Li et al. [242] first time proposed a Patch Transformer based architecture to automate the multi-tagging whole slide images process. They incorporated attention mechanism to extract global level features from the patches of images. They employed multi-tag attention module to build tag on the basis of weight. Extensive experiments demonstrated that their model outperformed on 4920 WSI for macro and micro F1 metric.

#### 6.5.7. OCT or Fundus Images

Diabetes damages the retina and causes diabetic retinopathy. This disease can lead to vision loss, therefore, early detection is crucial. Various deep learning approaches have automated the recognition of diabetic retinopathy grading. However, Wu et al. [158] proposed vision transformer based architecture to assist the ophthalmologist in recognising the diabetic retinopathy grade. They divided the fundus images into patches, flatten them to generate sequence, and then converted them into linear and positional embeddings. To generate the final representations, they fed the positional embeddings into multi-head attention layers. Their model outperformed the CNN-based architecture with an accuracy of 91.4%.

The table 8 given below summarized the performance gain by the reviewed articles of the miscellaneous category.

Table 8: A list of datasets and performance measures adopted by researchers for Miscellaneous Application in ViT

| Modality | Publication | Dataset | Performance Measures |
|---|---|---|---|
| CT Scans | Wang et al. [6] | NIH-AAPM-Mayo Clinic LDCT Grand Challenge [119] | RMSE = 8.7681 |
| | | | SSIM = 0.9144 |
| | Zhang et al. [233] | NIH-AAPM-Mayo Clinic LDCT Grand Challenge[119] | RMSE = 21.199 $\pm$ 2.054 |
| | | | SSIM = 0.933 $\pm$ 0.012 |
| | | | VIF = 0.144 $\pm$ 0.025 |
| PET Scans | Zhang et al.[234] | uPMR790 | RMSE = 0.0447 |
| | | | PSNR (db) = 27.5321 |
| | | | SSIM = 0.9291 |
| | | | PCC = 0.9899 |
| | Luo et al. [235] | Clinical dataset which includes eight normal control (NC) subjects. | PSNR = 24.818 |
| | | | SSIM = 0.986 |
| | | | NMSE = 0.0212 |
| | | | PSNR = 25.249 |
| | | Clinical dataset which includes eight mild cognitive impairment (MCI) subjects. | SSIM = 0.987 |
| | | | NMSE = 0.0231 |
| Endoscopy Images | | DIVerse 2K resolution high quality (DIV2K) images dataset [121] | |
| | | Set5 | PSNR (db) = 31.94 |
| | | | SSIM = 0.8935 |
| | | Set14 | PSNR (db) = 28.11 |
| | | | SSIM = 0.7842 |
| | | B100 | PSNR (db) = 27.54 |
| | | | SSIM = 0.7464 |
| | | Urban100 | PSNR (db) = 25.87 |
| | | | SSIM = 0.7844 |
| | | Manga109 | PSNR (db) = 30.09 |
| | | | SSIM = 0.9077 |
| MRI Scans | Wang et al. [236] | Kirby21 dataset (KKI01 to KKI05) [120] | PSNR = 40.19 ± 2.04 |
| | | | SSIM= 0.9882 ± 0.0034 |
| | Tang et al. [238] | T1-weighted images (of size 182×218×182) of 102 drug-addicts and 10 healthy volunteers. | Dice = 0.91 |
| | | | HD = 2.68 |
| | | | ASD = 0.59 |
| | Kokmaz et al. [3] | IXI dataset | |
| | | T1, R=4 | PSNR = 32.55 ± 1.77 |
| | | | SSIM (%) = 94.58 ± 0.82 |
| | | T1, R=8 | PSNR = 30.28 ± 1.68 |
| | | | SSIM (%) = 91.64 ± 1.42 |
| | | T2, R=4 | PSNR = 32.71 ± 0.73 |
| | | | SSIM (%) = 87.66 ± 1.67 |
| | | T2, R=8 | PSNR = 29.90 ± 0.70 |
| | | | SSIM (%) = 84.03 ± 1.89 |
| | Hu et al. [237] | IXI dataset | |
| | | T1 | PSNR = 34.91 $\pm$ 1.00 |
| | | | SSIM = 0.895 $\pm$ 0.07 |
| | | T2 | PSNR = 35.34 $\pm$ 0.95 |
| | | | SSIM (%) = 0.895 $\pm$ 0.07 |
| fMRI Scans | Zhang et al. [59] | fMRI experiments(the visual stimulus consisted of 2750 natural images from ImageNet data) | CIDEr = 0.741 |
| | | | ROUGE-L= 0.2009 |
| | | | BLEU= 0.186 |
| Fluorescence Microscopy | Gomariz et al. [240] | 3D confocal microscopy datasets | F1 Score = 0.912 $\pm$ 3.9 |
| Fundus Images | Wu et al. [158] | Diabetic Retinopathy Detection dataset | Accuracy (%) = 91.4 |
| | | | Specificity = 0.977 |
| | | | Precision = 0.928 |
| | | | Sensitivity = 0.926 |
| | | | Quadratic weighted kappa score = 0.935 |
| | | | AUC = 0.986 |
| WSI | Li et al. [242] | 4,920 WSIs provided by a histopathology service company | Mcro F1 = 0.910 |
| | | | Micro F1 = 0.944 |

## 7. Discussion and Conclusion

Vision Transformers (ViT) are now one of the hottest topics in the discipline of computer vision because of its exemplary performance and tremendous potential compared with CNNs. Although CNNs are matured enough for the development of applications that can ensure an efficient and accurate diagnosis. However, in the medical field, where an inaccurate output might endanger lives, the concept of attention in vision transformers has paved its way for more precise outcomes. Since ViT models assess the global context of the image along with the interpretability through the attention module, their performance is more precise than CNNs. A variety of approaches have been proposed in recent years, as outlined and summarized in this review, to explore and utilize the competency of vision transformers. These approaches showed excellent performance on a wide range of visual recognition tasks, including classification, lesions detection, anatomical structure segmentation, and clinical report generation. Nevertheless, the real potential of transformers for computer vision has yet to be fully explored, which means that significant challenges are still there to be resolved. The following section envisages these challenges as well as provides insights on future prospects.

### 7.1. Current Trends and Open Challenges

Although several transformer-based models have been proposed by researchers to address visual recognition tasks, these methods are simply the initial steps in this field and there is still considerable room for improvement. For example, the transformer architecture in ViT [19], is based on the standard transformer for NLP [16], although an enhanced version particularly tailored for CV needs to be explored. Furthermore, it is necessary to employ transformers to other medical domains such as orthodotics, medical report generation, and phenotype extraction etc., to discover the power of vision transformers.

In this multidisciplinary study, we provide a comprehensive view of how vision transformers employed medical imaging modalities in visual recognition tasks for assisted diagnosis. We have reviewed the research articles selected from top tier journals and conferences in which researchers have proposed excellent frameworks that employed vision transformers to accelerate the efficiency and performance of already proposed CNN based models. The Figure 5 is a clear evidence that vision transformers are widely employed in classification and segmentation related visual recognition tasks whereas their significance in registration related tasks have not been greatly explored. The miscellaneous studies in this review include the work on enhancing the resolutions for better outcomes, denoising of CT scans with a low dosage of X-ray radiation, image registration etc. Nevertheless, the undertaken studies are based on discrete applications, therefore, they are not comparable. The field of research is quite open at this stage as baselines are available. Then, report generation covered only 7% which indicates an area to research upon as its applications can assist both doctors and patients.

Similarly, in the imaging modalities, it can be observed in Figure 6 that 61% of the papers found, are related to X-rays, and CT scans. The vision transformer model was first proposed in the year 2020, and most of the corresponding literature available was related to the diagnosis of coronavirus as it was prevailing in the same year. The transformer models achieved admirable results in regards to COVID-19, however, other domains especially digital pathology, retinal, breast, etc., require attention as not much literature is available. Also, the literature related to COVID-19 is either classifying or detecting the virus, thus, no studies were found that were performing segmentation. Even though MRI produces images with better resolution than CT scans, it covered only 13% of the total modalities. The papers are mostly used for image segmentation and have produced respectable outcomes. These outcomes are based on a few studies targeting distinct diseases, hence, there is much more to explore.

In addition, it was observed in clinical report generation's review that among various modalities, the studies related to x-ray images were only available. Thus, it means that the data required for report generation is only present in correspondence to x-ray images. As X-rays are a cost-efficient method for examining various parts of the body, it is comprehensible that due to more reports and X-ray images available, it was possible to collect enough data for automatic report generation. Since it assists doctors in writing medical reports, the collection of corresponding images and text should be collected to implement automated report generations in other domains.

Next, one of the major limitations regarding medical datasets is that not enough data is available to train a transformer model. These models require data in huge amounts to perform well and adapt to generalizability. Another observation is the non-availability of a benchmark as observed in the Figure 16. Since almost all the studies are based on different datasets, the performances of the models are not comparable. Even if the datasets are the same in some studies, they are either used for different purposes or are combined with other datasets to form one large repository. Now, as long as there is no standard dataset for experimenting with different proposed models, research can not take its step forwards towards development using vision transformers.

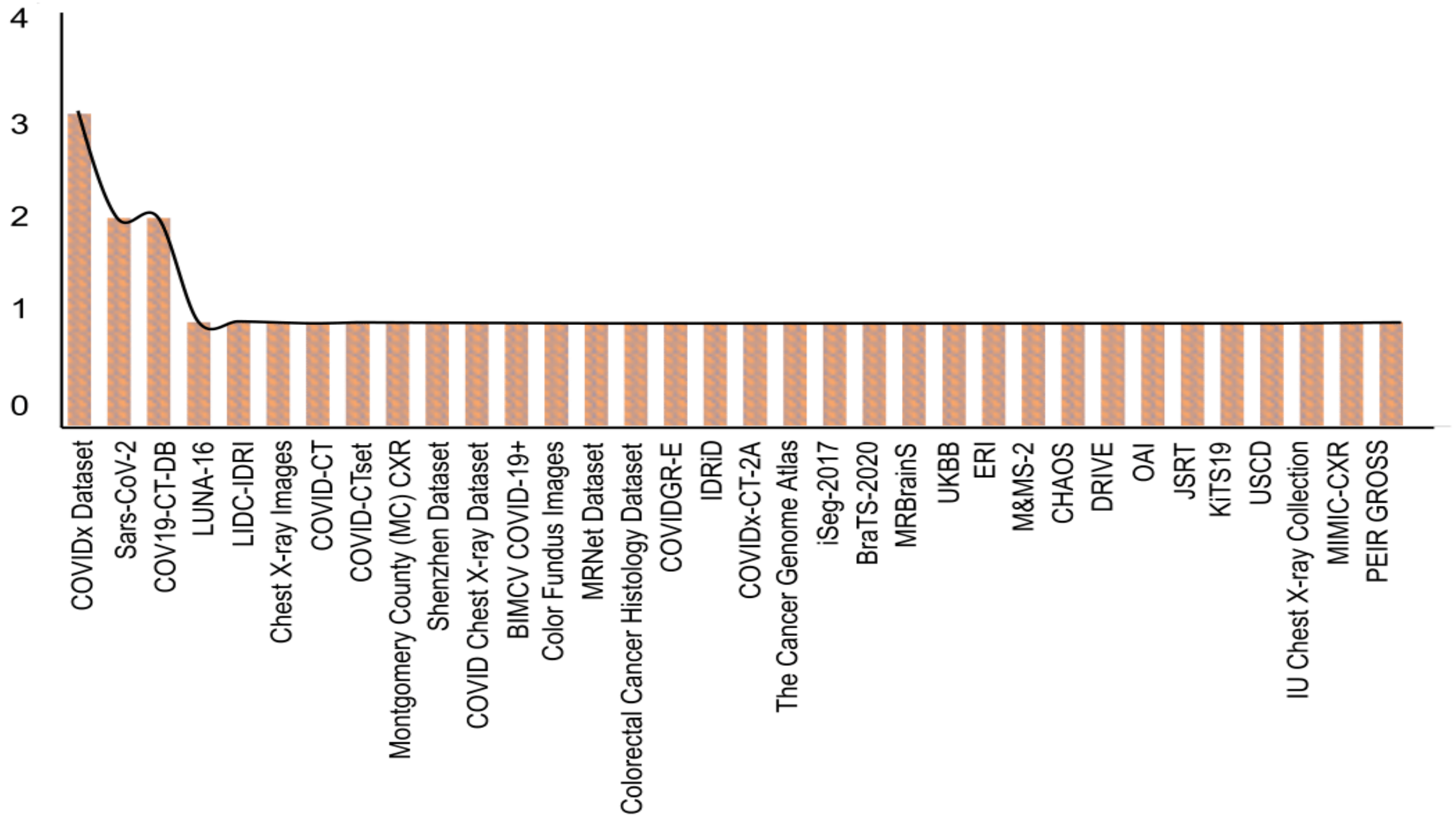

**Figure 16:** Long-tail graphical representation of datasets employed in literature.

## 7.2. Future Prospects:

In order to propel the development of vision transformers in medical domain, we propose various significant future directions. One direction is that, several papers in this study have done a comparative analysis of CNNs and ViT, thus, it can be inferred that transformer models have given better results in terms of efficiency, computational cost, and accuracy. Since these models have achieve such outcomes where researchers are talking about CNN's being replaceable, vision transformers require more research and implementation as they are unraveling a path towards more resource-efficient models.

The paper demonstrates the significance of using vision transformers on medical images. The future directions of research involve working with more heterogeneous data sets, and more extensive comparisons with other models to give validity to the proposed transformer models. Next, the studies, may they be related to any category or modality, are using different datasets, hence there cannot be a comparative analysis in terms of the proposed vision transformer models. Therefore, we should work on creating a benchmark dataset.

Similarly, new knowledge emerging in cancer biology and deep learning enabled us to step into this rapidly evolving domain. Genomic profiling is prevalent; however, it is crucial to correlate cancer genomic and phenotypic data to fully understand cancer behavior. Cancer phenotype information includes tumor morphology (e.g., histopathologic diagnosis), laboratory results (e.g., gene amplification status), specific tumor behaviors (e.g., metastasis), and response to treatment (e.g., the effect of a chemotherapeutic agent on tumor volume). Vision transformers can also be applied to medical images in order extract phenotypic information in order to better diagnose cancer related disorders. Another direction is related to the limitation regarding coronavirus cases being detected and classified, but not being segmented. The usage of segmenting the virus can help determine the rate of spread of the virus, that is why it should be worked upon in the future. Furthermore, the literature for clinical report generation is related to X-rays images, and as discussed above most of the vision transformers used for X-ray images are being used to detect coronavirus. As per our knowledge, none of the literature is related to medical report generation for coronavirus, and considering the work that has been done in this regard, research in medical image captioning can propel towards the application side and assist the practitioners. Overall, since it covers only 7 % of the total tasks in this study, there is more to explore in terms of X-rays and other imaging modalities. Although vision transformers have achieved another milestone for improved and accurate diagnosis in the medical domain, there is still room for improvement in terms of resource efficiency. In other words, we still have to discover the undiscovered.

### 7.3. Conclusion

The paper envisages the summary analysis of vision transformer models used in the medical domain. This study will serve researchers from multidisciplinary backgrounds. The visual recognition tasks considered in this paper include; classification, detection, segmentation, and clinical report generations. Other than that, some miscellaneous tasks were also included such as image registration, image reconstruction, etc. Furthermore, medical imaging modalities such as X-rays, CT Scans, MRIs, etc. were used as input to recognize medical imaging through various models. Our goal is to present the study in a way that apprehends the status and future directions of vision transformers. Hence, we structured the information of datasets, categories, modalities, and their results in a tabular form which will assist researchers to move forward in the medical field conveniently.